\documentclass[a4paper,11pt]{article}
\usepackage{jheppub}

\usepackage{graphicx}
\usepackage{svg}

\usepackage{amsmath}
\usepackage{mathtools}
\usepackage{slashed}
\usepackage{bbm}

\numberwithin{equation}{section}

\usepackage[english]{babel}
\usepackage[utf8]{inputenc}
\usepackage[T1]{fontenc}

\usepackage{comment}
\usepackage{savesym}
\savesymbol{Cross}
\usepackage{marvosym}
\restoresymbol{MVS}{Cross}

\usepackage{float}
\usepackage{amsthm}
\usepackage{amssymb}
\usepackage{amsbsy}
\usepackage{wrapfig}
\usepackage{bm}
\usepackage{enumerate}
\usepackage{bbding}
\usepackage{physics}
\usepackage{slashed}
\usepackage[clock]{ifsym}

\def \be {\begin{equation}}
\def \ee {\end{equation}}
\def \dd  {{\rm d}}

\newcommand{\hodge}{{\star}}

\usepackage{exercise}

\usepackage{media9}

\arxivnumber{2410.02616} 

\title{\boldmath Supergravity solutions for D$p$-D$(6-p)$ bound states: from $p=7$ to $p=-1$}

\author[a]{Sébastien Reymond,}
\author[b]{Mario Trigiante,}
\author[a]{Thomas Van Riet}

\affiliation[a]{
Instituut voor Theoretische Fysica, K.U. Leuven,\\
Celestijnenlaan 200D, B-3001 Leuven, Belgium}
\affiliation[b]{Department of Applied Science and Technology, Politecnico di Torino,\\
Corso Duca degli Abruzzi 24, I-10129 Turin, Italy and INFN, Sezione di Torino, Italy}

\emailAdd{sebastien.reymond@kuleuven.be}
\emailAdd{mario.trigiante@gmail.com}
\emailAdd{thomas.vanriet@kuleuven.be}

\abstract{Near horizon geometries of D$p$-branes with $p\neq 3$ are singular with a running dilaton. Bound states of D$p$ branes with their magnetic cousins, D$(6-p)$ branes, can stabilise the dilaton such that an AdS factor might appear in the near horizon region, potentially leading to a chain of AdS vacua of the form $AdS_{p+2}\times S^{p+2} \times \mathbb{T}^{6-2p}$.  The solutions with $p=-1, 1, 3$ are supersymmetric with the cases $p=1, 3$ being well-known examples already. We construct explicit (partially smeared) brane bound state solutions for all such configurations. The D2-D4 and D$(-1)$-D7 cases are entirely novel, but they do not have a near-horizon AdS geometry.  The two novel classes of solutions feature ghost branes (negative tension branes), and we suggest they are physical for the D$(-1)$-D7 solutions but unphysical for the D2-D4 solutions. The bound state of a D$(-1)$ and a D7 brane in supergravity was only hinted upon recently in \cite{Aguilar-Gutierrez:2022kvk}. We correct the solution here in order to preserve supersymmetry, and find that the dilaton can indeed be stabilized. This points to a possible dual matrix theory, generalizing the IKKT matrix model to allow for conformal invariance.}

\begin{document}
\maketitle
\flushbottom

\section{Introduction}
Supergravity $p$-brane solutions have been pivotal in our understanding of string theory and holography. Yet, many basic questions about supergravity $p$-branes remain unanswered. For instance, what are the solutions corresponding to bound states of branes? Such bound states are typically known when they preserve supersymmetry, although these solutions may be incomplete since the branes are often smeared\footnote{A smeared brane has a uniform charge distribution along some direction(s).} over some directions \cite{Tseytlin:1996bh,Bergshoeff:1996rn}. Consider for instance the well-studied D1-D5 bound state, where the D1 extends along the D5 worldvolume:
\begin{equation}
    \begin{aligned}
D1 \quad  & \times \times - - - - - - - - \\
D5\quad &  \times \times  - - - - \times \times \times \times \\
\end{aligned}
\end{equation}
where a cross denotes a worldvolume direction and a bar a transversal direction.  The metric for the known solution, in 10d string frame, is given by
\begin{equation}
ds^2_{10}= \frac{1}{\sqrt{H_1H_5}}\left(-dx^2_0 + dx^2_1\right) + \sqrt{H_1H_5}\left(dr^2 + r^2d\Omega_3^2\right) + \sqrt{\frac{H_1}{H_5}}\left(dx^2_6 + dx^2_7+ dx^2_8+dx^2_9\right)\,.
\end{equation}
Here $d\Omega_3^2$ denotes the metric on the normalised round 3-sphere. The functions $H_1$ and $H_5$ are harmonic
\begin{equation}
H_{1,5}(r) =1+\frac{|Q_{1,5}|}{r^2}\,,
\end{equation}
with the numbers $Q_1, Q_5$ being proportional to the brane charges. If $Q_1=0$ one obtains the D$5$  solution and if $Q_5=0$ we find the D$1$ solution smeared over the directions the $D5$ did not share with the D$1$. This is common for most of the known solutions for BPS bound states \cite{Smith:2002wn, Bardzell:2024anh}. Nonetheless, string theory informs us that a supersymmetric D$m-$Dn bound state can exist when the number of mixed boundary conditions (Dirichlet-Neumann) for the string equals a multiple of $4$. Clearly this is the case for the D$1-$D5 bound state. Another example would the following intersection of D3 branes:
\begin{equation}
    \begin{aligned}
D3 \,\quad  & \times \times \times \times - - - - - - \\
D3'\quad &  \times \times - - \times \times - - - - \\
\end{aligned}
\end{equation}
The solution for this bound state can similarly be written by inserting the harmonics in the metric in the right places, but the $D3$ stack will be smeared over the directions $4$ and $5$ and the $D3'$ stack over the directions $2$ and $3$.\footnote{T-duality along a spatial direction changes a cross for a bar. After a T-duality along the directions $2$ and $3$ we obtain the D1-D5 bound state with the D1 smeared over $2, 3, 4$ and $5$. T-duality preserves the number of directions with mixed boundary conditions.} Surprisingly, no solutions are known for which this smearing is absent, even though (from a string theory perspective) it is believed that localised solutions must exist. \\

In this paper, we are interested in bound states of the form D$p$-D$(6-p)$ for $p=-1,0,1,2,3$ since such solutions sometimes allow for smooth horizons potentially leading to AdS/CFT dual pairs. This is known to be the case for $p=1,3$. To see this, note that D$p$ branes are electrically charged under a $F_{p+2}$ field strength, and D$(6-p)$ branes are charged magnetically under the same field strength. Now, consider the dilaton equation in 10d Einstein frame:\footnote{Here we use the notation $F_{k}^2\equiv F_{\mu_1\dots\mu_k} F^{\mu_1\dots\mu_k}$.}
\begin{equation}
\nabla \partial \phi = \tfrac{3-p}{(p+2)!4}e^{\frac{(3-p)}{2}\phi}F_{p+2}^2+\ldots\,.
\end{equation}
Clearly, any electric charge leads to a non-zero and negative $F^2$ and thus a dilaton gradient, which diverges near the would-be horizon. Magnetic charges create a positive $F^2$ on the right hand side of the dilaton equation, and again leads to a dilaton gradient. The presence of both electric and magnetic charges can potentially make $F^2$ vanish, implying a constant dilaton solution. This happens for the D1-D5 system (or the D3 brane since there is no dilaton coupling). A zero $F^2$ at the horizon is then a consequence of the fluxes being (anti)-self dual in the directions along which the D1 brane is not smeared: the near horizon is $AdS_3\times S^3\times \mathbb{R}^4$, or equivalently $AdS_3\times S^3\times \mathbb{T}^4$. There is a physical motivation for choosing the torus, as the smearing of the $D1$ branes along the $\mathbb{T}^4$ directions can then be interpreted as a standard Kaluza-Klein coarse-graining procedure. The resulting solution can be thought of as a dyonic string solution in six dimensions. \\

It was pointed out in \cite{Aguilar-Gutierrez:2022kvk} that similar reasoning can apply to more general $p$-branes. For even $p$ we can have the D0-D6 bound states
\begin{equation}
\begin{aligned}
D0\, &\times - - - - - - - - - \\ 
D6\, &\times  - - - \times \times \times \times \times \times \\
\end{aligned}
\end{equation}
 or D2-D4 bound states
\begin{equation}
\begin{aligned}
D2 \times \times \times - - - - - - -\\ 
D4 \times  \times \times  - - - - - \times \times \\
\end{aligned}\,.
\end{equation}
For the D0-D6 system, the D0s are smeared over the $\mathbb{T}^6$, resulting in a dyonic black hole (particle) in 4d with horizon $AdS_2\times S^2$. Similarly, the D2s are smeared over a $\mathbb{T}^2$ along which the D4 branes extend. Despite naive expectations, we will show that there is no AdS horizon in this case. For odd values of $p$ we have the well-known 1/2 BPS dyonic D3 solution in IIB with its fully BPS near horizon $AdS_5\times S^5$. For $p=1$ we have the dyonic strings in 6d and for $p=-1$ one would expect dyonic instantons in 2d with an ``$AdS_1\times S^1$'' horizon. The latter was discussed in \cite{Aguilar-Gutierrez:2022kvk} and will be revised here. \\

The goal of this paper is to provide solutions for all cases not (or only partially) studied earlier in the literature: D0-D6, D2-D4 and D$(-1)$-D7. The motivations behind this goal are twofold:
\begin{enumerate}
    \item Extend our understanding of $p$-brane bound states, especially in non-SUSY cases. For example, the D2-D4 solution in this paper is entirely new. The D$0-$D6 was implicitly known since it is the 10d lift of the dyonic Kaluza-Klein black hole. The SUSY D$(-1)$-D7 solution presented here is also new and differs from the proposal in \cite{Aguilar-Gutierrez:2022kvk}.
    \item Holography: brane near horizons provide the decoupling limits for which one can argue for holographic dual pairs. Non-SUSY backgrounds tend to be at best meta-stable \cite{Ooguri:2016pdq}, making the definition of a holographic dual unclear \cite{Maldacena:1998uz}. For $p=-1$, the (SUSY) dual has been conjectured in \cite{Aguilar-Gutierrez:2022kvk} to be the matrix model studied in \cite{Billo:2021xzh}. If so, this constitutes a holographic pair where spacetime is emergent from matrices alone. Note that without adding the magnetic charges the duals are non-conformal. For $D0$ branes we have the conjectured BFSS quantum mechanics \cite{Banks:1996vh} and for D$(-1)$ branes the IKKT matrix model \cite{Ishibashi:1996xs}.
\end{enumerate}
In the next section we construct both the D0-D6 and D2-D4 bound state solutions by reducing the branes over their worldvolumes down to instantons, which in turn can be solved using basic group theory, as pioneered in \cite{Breitenlohner:1987dg} and \cite{Bergshoeff:2008be}.  To find the SUSY D$(-1)$-D7 bound state we instead solve the Killing spinor equations directly in section \ref{sec:D-1-D7}. We end with a discussion in section \ref{sec:discussion}. 

\section{Brane solutions from geodesics} \label{geodesics}
A particularly powerful method to obtain stationary brane solutions relies on the hidden symmetries that become manifest once a p-brane in $D$ dimensions is dimensionally reduced over its worldvolume to an instanton in a Euclidean theory in $D-p-1$ dimensions. This map between black holes and instantons through timelike reduction was first introduced in \cite{Breitenlohner:1987dg} and generalised to general $p$-branes in \cite{Bergshoeff:2008be}. The power of this method lies in the fact that the Euclidean equations of motion are such that the Einstein equations decouple almost completely from the equations of the matter fields, and the latter become the equations for a geodesic curve on some target space, often with more isometries than symmetries visible in the original theory. For instance, the Einstein-Maxwell dilaton theory in $D=4$ obtained from a $\mathbb{T}^6$ reduction of IIA supergravity enjoys a $SL(3, \mathbb{R})$ symmetry in 3D when compactified over time. The geodesic problem is then explicitly integrable and the lift of the instantons in 3d, described by the integrable geodesics, give the known black hole solutions in 4d.\\ 

Below we outline this procedure first for D$p$-D$(6-p)$ intersections that require no extra fields beyond $F_{p+2}$ (D0-D6, D1-D5) and then generalize to the case where the $B$-field is needed as well (D2-D4). Since the method is rather general, we will at first keep the notation with arbitrary $p$, even though we will only cover the cases $p=0,1, 3, 5 ,6$. 

\subsection{Brane bound state solutions without \texorpdfstring{$B$}{B}-field}

Consider the D$p$-D$(6-p)$ systems in 10d, which can be seen as $p$-brane solutions with magnetic and electric charges in $2p+4$ dimensions with self-dual $F_{p+2}$ field strengths at the near horizon region. The map from $10$ to $2p+4$ dimensions occurs through straightforward dimensional reduction over a $\mathbb{T}^{6-2p}$ and keeping only the overall volume modulus of the torus. We then end up with the following action in $2p+4$ spacetime dimensions
\begin{equation}
S =\int \sqrt{|g|}\left(\mathcal{R} -\tfrac{1}{2}(\partial \Phi)^2 - \tfrac{1}{2}\tfrac{1}{(p+2)!}e^{a\Phi}F_{p+2}^2 \right) \,, 
\end{equation}
where $\Phi$ is a particular linear combination of 10d dilaton and torus volume and $a$ is a specific number whose value matters for symmetry enhancement. From 10d supergravity we can deduce
\begin{equation}
    a^2 = 3-p\,.
\end{equation}
The dimensional reduction over $\mathbb{R}^{1,p}$ proceeds as follows
\begin{align}
& ds^2_{2p+4} = e^{2\alpha\varphi} ds^2_{p+3} + e^{2\beta\varphi} ds^2_{p+1} \,,\\
& \hat{C}_{p+1} = C_{p+1}+ \chi_E\epsilon_{p+1}\,,
\end{align}
where the hat indicates the form field in the higher dimension and  $\chi_E$ will become an axion in $p+3$ dimensions whose axion charge describes electric charge in the higher dimension. When we choose
\begin{equation}
 \alpha^2 =\frac{1}{4(p+1)}\,,\qquad \alpha=-\beta\,,   
\end{equation}
the dimensional reduction and truncation give the following Lagrangian density in $p+3$ Euclidean dimensions
\begin{equation} \label{instantonaction1}
\frac{\mathcal{L}}{\sqrt g} =\mathcal{R} -\tfrac{1}{2}(\partial \Phi)^2 - \tfrac{1}{2}(\partial \varphi)^2 + \tfrac{1}{2}e^{a\Phi +(p+1)(\alpha-\beta)\varphi}(\partial \chi_E)^2 - \tfrac{1}{2}\tfrac{1}{(p+2)!}e^{a\Phi +(p+1)(\beta-\alpha)\varphi}F_{p+2}^2 \,.
\end{equation}
The odd-sign axion kinetic term is caused by reducing over a space with 1 timelike dimension.  
In this $p+3$-dimensional Euclidean theory we can Hodge dualise the $F_{p+2}$ to a 1-form axion field-strength whose axion potential we denote $\chi_M$ since it describes magnetic charges in the higher dimensions. The resulting action is (up to boundary terms)
\begin{equation}
\frac{\mathcal{L}}{\sqrt{g}}=\mathcal{R} -\tfrac{1}{2}(\partial \Phi)^2 - \tfrac{1}{2}(\partial \varphi)^2 + \tfrac{1}{2}e^{a\Phi +(p+1)(\alpha-\beta)\varphi}(\partial \chi_E)^2 + \tfrac{1}{2}e^{-a\Phi +(p+1)(\alpha-\beta)\varphi}(\partial \chi_M)^2  \,. 
\end{equation}
This sigma model is not a symmetric coset, but it can be embedded into a symmetric coset. For instance for D0-D6, it can be embedded into $SL(3,\mathbb{R})/SO(2,1)$. 

If we look for null geodesics then the energy-momentum of the 4-scalar fields cancels out and the $p+3$-dimensional metric is flat:
\begin{equation}
 ds^2_3 =dr^2 + r^2d\Omega_{p+2}^2\,.   
\end{equation}
It is useful to work with a different radial coordinate 
\begin{equation}
r   \sim \tau^{-\frac{1}{p+2}}\,,
\end{equation}
since then $\tau$ is an affine coordinate on the geodesic curve traced out in the 4d target space. In other words, the equations of motion in that coordinate system are
\begin{align}
& \ddot{\phi}^i + \Gamma^i_{jk}\dot{\phi}^j\dot{\phi}^k =0\,,\\
& G_{ij}\dot{\phi}^i\dot{\phi}^j =0
\,.
\end{align}
We have used a notation in which $\phi^i=\left\{ \Phi, \varphi, \chi_M, \chi_E \right\}$ ($i=1,\ldots 4$) and $G_{ij}$ is the metric that appears in the kinetic term $\frac{1}{2}G_{ij}\dot{\phi}^i\dot{\phi}^j$, the $\Gamma$ are the corresponding Christoffel symbols and a dot is a derivative with respect to $\tau$.

One way to understand that for the correct values of the coefficient $a$ the sigma model can be embedded as a truncation of a symmetric coset, and hence must be integrable, comes from integrating out the axion momenta. When doing so the effective geodesic action is given by the classical mechanics system:
\begin{equation}
L =    -\tfrac{1}{2}(\dot\Phi)^2 - \tfrac{1}{2} (\dot\varphi)^2 - \tfrac{1}{2}e^{-a\Phi -(p+1)(\alpha-\beta)\varphi} Q_E^2 - \tfrac{1}{2} e^{a\Phi -(p+1)(\alpha-\beta)\varphi}Q_M^2  \,.
\end{equation}
This is a system of two generalised coordinates in a potential that is the sum of exponentials. These systems are known to be integrable when they are of the ``Toda-molecule" kind \cite{Ferreira:1984bi}. This happens when the vectors $\vec{\alpha}_i$ of exponentials in the potential $e^{\vec{\alpha}_i\cdot\vec{q}}$ obey that the following matrix
\begin{equation}
A_{ij} =2\frac{\vec{\alpha_i}\cdot\vec{\alpha_j}}{||\vec{\alpha}_i||^2} \,.
\end{equation}
corresponds to the Cartan matrix of a semi-simple Lie algebra. In our case we have
\begin{align}
&\vec{\alpha}_1 = (-a, -2(p+1)\alpha)\,,\\
&\vec{\alpha}_2 = (+a, -2(p+1)\alpha)\,.
\end{align}
If we combine our expressions for $a$ and $\alpha$ we find:
\begin{align}
&\vec{\alpha}_1\cdot \vec{\alpha}_2 = 2(p-1)\,.\\
&\vec{\alpha}_1\cdot \vec{\alpha}_1 = \vec{\alpha}_2\cdot \vec{\alpha}_2=4\,.
\end{align}
Hence:
\begin{align}
& D3-D3: A = 2,\quad  A_1 \quad (\text{just one exponential})\\
& D2-D4: A =\begin{pmatrix} 2 & 1 \\
1 & 2 \\
\end{pmatrix}\\
& D1-D5: A =\begin{pmatrix} 2 & 0 \\
0 & 2 \\
\end{pmatrix}\quad \text{so} \quad A_1\oplus A_1 \\
& D0-D6: A =\begin{pmatrix} 2 & -1 \\
-1 & 2 \\
\end{pmatrix} \quad \text{so}\quad A_2=SL(3)\,.
\end{align}
Only the D2-D4 is not a Cartan matrix. Yet, the equations of motion for the D2-D4 system must be integrable since it can be obtained from a consistent truncation of IIA on $\mathbb{T}^2 \times \mathbb{R}^{1,2}$ which is 5d Euclidean supergravity with coset space $E_{6(6)}/H$, with $H$ some non-compact maximal subgroup. All such geodesics are integrable.\footnote{It has been claimed that indeed the integrability of Toda systems exactly comes from the embedding into a larger geodesic system \cite{Ferreira:1984bi}. The other way around, one can integrate out shift symmetric directions of an integrable geodesic motion to obtain an integrable system with potential.} This can be understood as a consequence of needing to keep the $B$-field for consistency of the truncation. We turn to this after we showcase the D0-D6 solution.

\subsection{The \texorpdfstring{D$0-$D6}{D0-D6} bound state solution}
\label{sect:D0_D6_solution}

By lifting the so-named dyonic Kaluza-Klein black hole solution in 4d, which can be constructed using the geodesic approach outlined above, one finds the D0-D6 brane intersection without B-field\footnote{With B-field it can be supersymmetric \cite{Witten:2000mf}.} This was for instance done in \cite{Chemissany:2010zp}. For more information on the stringy physics of this set-up we refer the reader to \cite{Witten:2000mf, Taylor:1997ay, Sheinblatt:1997nt, Lee:2008ha}.

The KK black hole is a solution to the following theory of a scalar field $s$ and a one-form field $A_1$ with two-form field strength $F_2 =d A_1$ coupled to gravity in 4d:
\begin{equation}\label{Einsteindilatonmaxwell}
S = \int\sqrt{-g}\left(R - \tfrac{1}{2}(\partial s)^2 -\tfrac{1}{4}e^{\sqrt3 s }F_2^2 \right)\,.
\end{equation}
For this particular coupling between the scalar and the vector, the system of differential equations for spherically symmetric and static solutions is Liouville-integrable such that group theory techniques provide the solutions, see e.g. \cite{Breitenlohner:1987dg, Chemissany:2010zp}. We review the extremal solution in Appendix \ref{app:dilatonic_BH_uplift}.  Interestingly, it is also this $\sqrt3$-coupling that one obtains by dimensionally reducing IIA supergravity on a 6-torus. The electric charge of the 4d black holes then lift to D0 charges and the magnetic charges in 4d lift to D6 charges. Hence, uplifting the dyonic extremal solution leads to a D0-D6 bound state. All of this is reviewed in Appendix \ref{app:dilatonic_BH_uplift} and we simply present the solution here in string frame
\begin{align}
    \dd s^2_{10,s} &= g_s^{1/2}\left(
    - H_0^{-\tfrac{1}{2}}H_6^{-\tfrac{1}{2}} 
    \dd t^2 + 
    H_0^{\tfrac{1}{2}}
    H_6^{\tfrac{1}{2}}
    \left(\dd r^2 + r^2 \dd\Omega_2^2\right)+
    H_0^{\tfrac{1}{2}}H_6^{-\tfrac{1}{2}} \delta_{ij}\dd \theta^i \dd \theta^j\right) \, ,\\
    e^{\phi} &= g_s\left(\frac{H_0}{H_6}\right)^{\tfrac{3}{4}} \, ,\\
    F_2 &= Q_6 \dd \Omega_2 + g_s^{-3/2} Q_0 \frac{H_0^{-2}H_6}{r^2} \dd t\wedge \dd r \, ,
\end{align}
where $\theta^i$ are coordinates on flat space, whether a  6-torus, $\mathbb{R}^6$ or something else. The functions $H_0$ and $H_6$ are\footnote{The notation $Q^{2/3}$ should be interpreted as $(Q^2)^{1/3}$, so that the solution is valid for any signs of the charges.}
\begin{equation}
    H_0(r)  =  1 + g_s^{-1/2}Q_0^{2/3} G(r)\, ,\qquad
    H_6(r)  =  1 + g_s^{1/2}Q_6^{2/3} G(r)\,,
\end{equation}
with
\begin{equation}
    G(r) \equiv \frac{1}{r}
        \sqrt{g_s^{-1/2}Q_0^{2/3} + g_s^{1/2}Q_6^{2/3}} +  \frac{Q_0^{2/3}Q_6^{2/3}}{2r^2}\, .
\end{equation}
The effect of combining both D0 and D6 charges is to introduce subleading $1/r^2$ terms in what used to be harmonic functions on the space transversal to both branes. One readily verifies that the solution has an AdS$_2\times$S$^2$ near horizon.

In what follows we will analyse one consequence of these terms and these are so-called \emph{brane-jet instabilities} \cite{Bena:2020xxb} which verify the Swampland conjecture that all non-SUSY AdS vacua must have some form of instability \cite{Ooguri:2016pdq}. Consider for instance the action for a probe (anti-)D0 brane:
\begin{align}
    S_{D_0} &= -\mu_0 e^{-\phi} \int \dd\tau \, \sqrt{-\gamma} \pm \mu_0 \int C_{1}\,,
\end{align}
where $\gamma$ is the background metric pulled back to the brane worldvolume, $C_1$ is the background gauge potential and $\mu_0$ is the (absolute value of the) charge. A local expression for the t-component of the $C_1$-field is:
\begin{equation}
(C_1)_t = g_s^{-3/2}Q_0 \frac{1}{r^2 H_0(r)} \left(\frac{1}{2}g_s^{1/2}Q_6^{2/3}\sqrt{g_s^{-1/2}Q_0^{2/3}+g_s^{1/2}Q_6^{2/3}} + r\right) + C_{\text{int}}\, ,
\end{equation}
where $C_{\text{int}}$ is an integration constant that will be irrelevant for our discussion. Hence the potential felt by the probe is
\begin{equation}
V \propto g_s^{-3/4}H_0^{-1}H_6^{1/2} + g_s^{-3/2}|Q_0| \frac{1}{r^2 H_0(r)} \left(\frac{1}{2}g_s^{1/2}Q_6^{2/3}\sqrt{g_s^{-1/2}Q_0^{2/3}+g_s^{1/2}Q_6^{2/3}} + r\right)\, ,
\end{equation}
where we fixed the $\mp$ sign based on the sign of $Q_0$ to get the expected electromagnetic repulsion for the probe D0. It is easy to verify that the potential is positive everywhere and monotonically decreasing.  Therefore, the probe D0 brane is pushed to infinity: there is indeed a brane jet instability.

\subsection{The \texorpdfstring{D$2-$D$4$}{D2-D4} bound state requires the B-field}\label{sec:D2-D4}
We consider the brane intersection described by the following table:
\begin{equation*}
    \begin{array}{l|lllllllll}
\mathrm{x} & \mathrm{x} & \mathrm{x} & - & - & - & - & - & - & - \\
\mathrm{x} & \mathrm{x} & \mathrm{x} & \mathrm{x} & \mathrm{x} & - & - & - & - & -
\end{array}
\end{equation*}
The would-be supergravity solution could be expected to be a solution to the following truncation of type IIA supergravity:
\begin{align}
    S_{10} = \int \dd^{10}x \sqrt{-g} \left(\mathcal{R}-\frac{1}{2}(\partial \phi)^2 - \frac{1}{4!2}e^{\frac{1}{2} \phi}F_4^2\right)\, .\label{eq:D2_D4_10d_action_without_B_field}
\end{align}
However, this cannot be correct since in the absence of $F_2$ and $F_0$ (and all fermions) we have the following form equations (in Einstein frame):
\begin{align}
& d \left( e^{\phi/2}\star F_4\right) = -H_3 \wedge F_4\,, \\   
& d \left( e^{-\phi} \star H_3\right) = \frac{1}{2}F_4\wedge F_4\,.
\end{align}
Since our configuration of interest would involve both electric and magnetic $F_4$ charges we expect $F_4\wedge F_4$ to be non-vanishing, meaning that the $H_3$ field cannot be truncated. Instead, the minimal truncation of the full IIA action we need is
\begin{align}
    S_{10}  = \int  \left(  d^{10}x\,\sqrt{-g} \left(\mathcal{R}-\frac{1}{2}(\partial \phi)^2 - \frac{1}{4!2}e^{\frac{1}{2} \phi}F_4^2- \frac{1}{3!2}e^{-\phi}H_3^2\right) +\frac{1}{2}F_4 \wedge F_4\wedge B_2\right)\, . \label{eq:D2_D4_10d_action}
\end{align}
For reasons explained earlier we reduce this over $\mathbb{T}^2$ using the following truncation
\begin{align}
& ds^2_{10} = e^{2\tilde{\alpha}\tilde{\varphi}}ds^2_{8} + e^{2\tilde{\beta}\tilde{\varphi}}(d\theta_1^2+d\theta_2^2)\,,\\
& B_2 = b d\theta_1\wedge d\theta_2\,,\\
& \hat{C}_3 = C_3\,.
\end{align}
The last identity means we do not consider legs of $C_3$ along the two torus directions $\theta_{1.2}$. The scalar $\tilde{\varphi}$ is the torus volume modulus and the numbers $\tilde{\alpha},\tilde{\beta}$ are chosen to get canonical normalization for the Einstein-Hilbert term and the volume modulus in 8d:
\begin{equation}
\tilde{\alpha}^2 = \frac{1}{48}\,,\qquad \tilde{\beta}=-3\tilde{\alpha}.     
\end{equation}
The 8d theory contains the metric, the volume scalar $\tilde{\varphi}$ and axion $b$ and a 3-form $C_3$:
\begin{align}
   S_8  = \int  d^{8}x\,\sqrt{-g}\left(\mathcal{R}-\frac{1}{2}(\partial \phi)^2 - \frac{1}{2}(\partial \tilde{\varphi})^2 - \frac{1}{2}e^{-\phi -4 \tilde{\beta}\tilde{\varphi}}(\partial b)^2  - \frac{1}{4!2}e^{\frac{1}{2}\phi +2\tilde{\beta}\tilde{\varphi}}F_4^2\right)  + \frac{1}{2} b F_4 \wedge F_4\,.
\end{align}
We notice that the only scalar direction that couples to the forms $b$ and $C_3$ is 
\begin{align}
\Phi \equiv \tfrac{1}{2}\phi +2\tilde{\beta}\tilde{\varphi}\,. \label{eq:D2D4_reduction_T2_field_redef_Phi}     
\end{align}
The orthogonal field direction will be truncated to some arbitrary constant $C$:
\begin{equation}
2\tilde{\beta}\phi -\tfrac{1}{2}\tilde{\varphi} = C\,. \label{eq:D2D4_reduction_T2_field_redef_C}      
\end{equation}
So our 8d theory becomes
\begin{align}\label{8d}
    S_8  = \int\left( d^{8}x\, \sqrt{-g}\left( \mathcal{R}-\frac{1}{2}(\partial \Phi)^2  - \frac{1}{2}e^{-2\Phi}(\partial b)^2  - \frac{1}{4!2}e^{\Phi}F_4^2\right)  + \frac{1}{2} b F_4 \wedge F_4\right)\, .
\end{align}

Our goal is to compute the extension of \eqref{instantonaction1} when $p=2$ that includes the $B$-field. We do this by reducing the Lagrangian density \eqref{8d} over $\mathbb{R}^{1,2}$ as before but keeping now also the axion $b$. We find
\begin{align}
\frac{\mathcal{L}_5}{\sqrt{g}} =& \mathcal{R} -\tfrac{1}{2}(\partial \Phi)^2 -  \tfrac{1}{2}(\partial \varphi)^2 + \tfrac{1}{2}e^{\Phi +3(\alpha-\beta)\varphi}(\partial \chi_E)^2 - \tfrac{1}{2}\tfrac{1}{4!}e^{\Phi +3(\beta-\alpha)\varphi}F_{4}^2 \nonumber\\
& -\tfrac{1}{2}e^{-2\Phi}(\partial b)^2 + \frac{b d\chi_E\wedge F_4}{\sqrt g}\,,
\end{align}
with $\alpha-\beta=\frac{1}{\sqrt3}$.
We then Hodge dualise $F_4$ in order to display the magnetic potential $\chi_M$ through:
\begin{equation}
F_4 = e^{-\Phi - 3(\beta-\alpha)\varphi}\star\left(d\chi_M + b d\chi_E\right)\,,
\end{equation}
and we find
\begin{equation} \label{L5}
\frac{\mathcal{L}_5}{\sqrt{g}} = \mathcal{R} -\tfrac{1}{2}(\partial \Phi)^2 -  \tfrac{1}{2}(\partial \varphi)^2 -\tfrac{1}{2}e^{-2\Phi}(\partial b)^2  + \tfrac{1}{2}e^{\Phi +\sqrt3\varphi}(\partial \chi_E)^2 
 +\tfrac{1}{2} e^{-\Phi +\sqrt3\varphi}(\partial \chi_M + b\partial\chi_E)^2 \,.    
\end{equation}
The sigma model thus obtained can be shown to be $SL(3,\mathbb{R})/SO(2,1)$. The concrete coset representative that gives this metric can be found as follows:
\begin{equation}
L = \exp{\chi_E E_{12}} \exp{\chi_M E_{13}} \exp{b E_{23}} \exp{ (-\tfrac{1}{4}\Phi-\tfrac{\sqrt3}{4}\varphi)H_1 + (\tfrac{\sqrt3
}{4}\Phi-\tfrac{1}{4}\varphi)H_2} 
\end{equation}
where the Cartan generators and positive step operators are given by
\begin{align}
& H_1 =\begin{pmatrix}
 1 & 0 & 0\\
 0 & -1 & 0\\
 0 & 0 & 0
\end{pmatrix}\,,\qquad  H_2 =\frac{1}{\sqrt3 }\begin{pmatrix}
 1 & 0 & 0\\
 0 & 1 & 0\\
 0 & 0 & -2
\end{pmatrix}\,,\nonumber\\
 & E_{12}= \begin{pmatrix}
 0 & 1 & 0\\
 0 & 0 & 0\\
 0 & 0 & 0
\end{pmatrix}\,,\qquad 
E_{13}= \begin{pmatrix}
 0 & 0 & 1\\
 0 & 0 & 0\\
 0 & 0 & 0
\end{pmatrix}\,,\qquad 
E_{23}= \begin{pmatrix}
 0 & 0 & 0\\
 0 & 0 & 1\\
 0 & 0 & 0
\end{pmatrix}\,. 
\end{align}
The sigma model can then be found from the metric
\begin{equation}
\text {Tr} (dM dM^{-1})\qquad \text{where}\quad M = L\eta L^T\,,\quad
\text{with}\quad \eta =\begin{pmatrix} 
    -1 & 0 & 0 \\
     0 & +1 & 0 \\
     0 & 0 & + 1
\end{pmatrix}    \,.
\end{equation}
Normalisations are set such that the Lagrangian density \eqref{L5} now becomes: 
\begin{equation}
\frac{\mathcal{L}_5}{\sqrt{g}} = \mathcal{R} + \tfrac{1}{4}\text{Tr}(\partial M \partial M^{-1}) \,.
\end{equation}
The coset element $L$ explicitly reads:
\begin{equation}
 L =
 \begin{pmatrix}
 e^{-\frac{1}{\sqrt3}\varphi}  &  e^{\frac{1}{2}\Phi + \frac{1}{2\sqrt3}\varphi}\chi_E & e^{-\frac{1}{2}\Phi +\frac{1}{2\sqrt3}\varphi}\left(b\chi_E +\chi_M\right) \\
 0 &  e^{\frac{1}{2}\Phi + \frac{1}{2\sqrt3}\varphi} & e^{-\frac{1}{2}\Phi +\frac{1}{2\sqrt3}}b\\
 0 & 0 & e^{-\frac{1}{2}\Phi +\frac{1}{2\sqrt3}}
\end{pmatrix}\,.
\end{equation}
Once the solutions to the geodesics are found, the uplift to 10d \emph{string frame} is given by
\begin{align}
ds^2_{10} =& e^{-\frac{1}{\sqrt3}\varphi -\frac{2C}{\sqrt3}}\left(-dt^2 + dx^2 + dy^2 \right) + e^{\frac{1}{\sqrt3}\varphi -\frac{2C}{\sqrt3}}\left(dr^2 + r^2d\Omega_4^2 \right) + e^{\Phi}\left(d\theta_1^2 + d\theta_2^2 \right)\,,\\
\phi =& \frac{1}{2}\Phi - \frac{\sqrt{3}}{2}C \,,\\
H_3 =& b'\,dr\wedge d\theta_1\wedge d\theta_2\,,\\ 
F_4 =& \chi_E'dr\wedge dt\wedge dx\wedge dy + e^{-\Phi+\sqrt3 \varphi}\left(\chi_M' + b\chi_E'\right) r^4 d\Omega_4\,.
\end{align}
Primes denote derivatives with respect to $r$ and the $r$-dependence of the quantities $\Phi, \varphi, b, \chi_M, \chi_E$ all comes from them being functions of the affine parameter $h(r)$\footnote{In particular $f'= (df/dh) (dh/dr) = -3 \dot{f} r^{-4}$.}.
Using the axion shift symmetries we deduce the two Noether charges, equal to the RR charges under $F_4$ ($Q_E$) and $F_6$ ($Q_M$):
\begin{align}
 &Q_M = e^{-\Phi+\sqrt3 \varphi}\left(\chi_M' + b\chi_E'\right) r^4\,, \\
 &Q_E = e^{\Phi+\sqrt3 \varphi} \chi_E' r^4 + b Q_M\,.
\end{align}
This allows us to rewrite the $F_4$ as
\begin{equation}
 F_4 =   \frac{(Q_E -b Q_M)}{r^4}e^{-\Phi -\sqrt3 \varphi} (dr\wedge dt\wedge dx\wedge dy)  +  Q_M d\Omega_4\,.
\end{equation}

\subsection{The \texorpdfstring{D$2-$D$4$}{D2-D4} null geodesics}
The affine parameter for the null geodesics is the radial harmonic $h(r)$ on 5d flat space 
\begin{equation}
h(r) = A + \frac{B}{r^3}\,,    \label{eq:affine_param_D2_D4}
\end{equation}
with $A$ and $B$ constants. We fix conventions such that $A=0, B=1$. The geodesics through the origin are solutions given by the exponential map, at least at the level of the symmetric coset matrix $M$:
\begin{equation}
M =  \eta \exp{Q h(r)}\,,   
\end{equation}
where $Q$ is a matrix inside the coset algebra:
\begin{equation}
Q^T = \eta Q \eta \,.    
\end{equation}
These are all geodesics through the origin. The expression can trivially be generalised. 

Null geodesics obey $\text{Tr}(Q^2)=0$ and we will search for all of them. The null condition is required by extremality of the corresponding \cite{Breitenlohner:1987dg, Bergshoeff:2008be} configuration. One way to see this, is that for lightlike geodesics we can immediately replace the radial harmonic for a harmonic with multiple centers and still have a solution to all equations of motion. This is identical to a no-force condition, which is the smoking gun of extremality. The matrix $Q$ is inside the Lie algebra $SL(3,\mathbb{R})$ and obeys 
\begin{equation}
Q^3+\frac{1}{2}\text{Tr}(Q^2)Q -{\rm det}(Q)\mathbbm{1} =0\,,     
\end{equation}
as a consequence of the Caley-Hamilton theorem for $3\times 3$ matrices that obey $\text{Tr}(Q)=0$. Lightlike geodesics require $\text{Tr}(Q^2)=0$ so we find 3 distinct options:  either the matrix is nilpotent of degree 2 ($Q^2=0$), degree 3 ($Q^3=0$) or it is not nilpotent and obeys $Q^3\propto \mathbbm{1}$. We have verified that the solution with nilpotent Q-matrix of order 2 correspond to either a stack of pure D2 branes or pure D4 branes. If we want both D2 and D4 charges at the same time we need to go beyond. Below we first present the solution for Q-matrices obeying $Q^3=0$ and then we discuss the solutions without nilpotent $Q$-matrix.

\subsubsection*{Solutions with $Q^3=0$}
 A general $Q$ matrix that vanishes at order $3$ is given by:
\begin{equation}
Q = \begin{pmatrix}
        -\frac{\alpha+\beta}{2} & -\frac{\beta}{2}\sqrt{\frac{\beta}{\beta-\alpha}} & \frac{\alpha}{2}\sqrt{\frac{\alpha}{\alpha-\beta}}\\
        \frac{\beta}{2}\sqrt{\frac{\beta}{\beta-\alpha}} & \frac{\beta}{2} & 0\\
        -\frac{\alpha}{2}\sqrt{\frac{\alpha}{\alpha-\beta}} & 0 & \frac{\alpha}{2}
    \end{pmatrix}\,.    
\end{equation}
This $Q$ matrix contains two constants, $\alpha$, $\beta$. If we put either to zero we find a matrix of nilpotency degree two, which indicates these numbers directly relate to D2 and D4 charges (denoted $Q_2, Q_4$). By comparing $M=\eta \exp\left(Qh\right)$ and $M = L \eta L^T$, one can extract the fields $b(h)$, $\varphi(h)$, $\Phi(h)$, and $\chi_{E/M}(h)$:
\begin{align}
    b(h) &= \frac{(-\alpha\beta)^{3/2}h^2}{h^2\alpha^2\beta-4h\alpha(\alpha-\beta)-8(\alpha-\beta)}\,,\\
    \chi_E(h) &= -\frac{(-\beta)^{3/2}}{\sqrt{\alpha-\beta}}\frac{h^2\alpha+4h}{h^2\alpha\beta+4h(\alpha+\beta)+8}\,,\\
    \chi_M(h) &= -\frac{\alpha^{3/2}}{\sqrt{\alpha-\beta}}\frac{h^2\beta+4h}{h^2\alpha\beta+4h(\alpha+\beta)+8}\,,\\
    e^{\Phi(h)} &= 2\sqrt{2}(\alpha-\beta)\frac{\sqrt{h^2\alpha\beta+4h(\alpha+\beta)+8}}
    {-h^2\alpha^2\beta+4h\alpha(\alpha-\beta)+8(\alpha-\beta)}\,,\\
    e^{\frac{1}{\sqrt{3}}\varphi(h)} &= \frac{1}{2}\left(\frac{1}{2}h^2\alpha\beta+2h(\alpha+\beta)+4\right)^{1/2}\,.
\end{align}
 The above solution is consistent for $\alpha \ge 0\ge \beta$, $\alpha\neq \beta$. Using these explicit expressions we find the following expressions for the $D2$ and $D4$-brane charges:
\begin{equation}\label{D2D4charge}
    Q_E = \frac{3}{2} \frac{(-\beta)^{3/2}}{\sqrt{\alpha-\beta}}\,,\qquad
    Q_M = \frac{3}{2} \frac{\alpha^{3/2}}{\sqrt{\alpha-\beta}}\,.
\end{equation}
Even though it seems that we have found our sought-for D2-D4 solution, the solution actually does not reduce to standard D2 or D4 brane solutions separately. To see this, we take the limit where either one of the charges vanishes. For instance, the D2 brane solution can be obtained by setting $Q_M=0$. In that case, we find the string frame metric, and field-strength
\begin{align}
    \dd s^2_{10} &= H_2^{-1/2}e^{-\frac{2C}{\sqrt3}}\left(-\dd t^2 + \dd x^2 + \dd y^2 \right) + H_2^{1/2}e^{-\frac{2C}{\sqrt3}}\left(\dd r^2 + r^2\dd \Omega_4^2 \right) + H_2^{1/2}\left(d\theta_1^2 + d\theta_2^2 \right)\\
    F_4 &= -\dd H_2^{-1} \wedge \dd t\wedge \dd x\wedge \dd y\,,
\end{align}
where we used
\begin{align}
    e^{\Phi(h)} &= e^{\frac{1}{\sqrt{3}}\varphi(h)} = \left(1-\frac{Q_E}{3r^3}\right)^{1/2} \equiv H_2^{1/2}\, .
\end{align}
Note the minus sign in the harmonic function: this describes a ghost brane for positive $Q_E$ and a D-brane for negative $Q_E$. When we instead keep the D4 charge, one finds the expected D4 solution with now
\begin{equation}
H_4=1+\frac{Q_M}{3r^3}\,.    
\end{equation}
But since the sign of $Q_E$ has to be equal to the sign of $Q_M$ (see above), this means that we always encounter one ghost brane. Indeed, one can easily verify that our lifted solution always becomes complex in the interior, hitting the typical singularity for a ghost brane when we include both charges described by our $Q^3=0$ solution. 

One possible resolution could be that the above $Q$-matrix is not the general $Q$-matrix that solves $Q^3=0$. Indeed, in Appendix \ref{app:Qcube} we explain there are disconnected branches of $Q$-matrices that differ in signs. Unfortunately all these matrices lead to solutions in which one brane source is ghost-like. For this reason, we move to the extremal solutions which have $Q^3 \sim \mathbbm{1}$ instead of nilpotent $Q$.

\subsubsection*{Solutions with $Q^3\neq 0$}
Upon fixing the action of  ${\rm SO}(2)\subset {\rm SO}(1,2)$, the $Q$ matrix can be written in the form:
\begin{equation}
  Q= \left(
\begin{array}{ccc}
 \frac{1}{2} (-\alpha -\beta ) & \frac{1}{2} \sqrt{-\frac{\beta ^3-8 \lambda ^3}{\alpha -\beta }} & \frac{1}{2} \sqrt{\frac{\alpha ^3-8 \lambda ^3}{\alpha -\beta }} \\
- \frac{1}{2} \sqrt{-\frac{\beta ^3-8 \lambda ^3}{\alpha -\beta }} & \frac{\beta }{2} & 0 \\
 -\frac{1}{2} \sqrt{\frac{\alpha ^3-8 \lambda ^3}{\alpha -\beta }} & 0 & \frac{\alpha }{2} \\
\end{array}
\right)\,,\label{matQ}
\end{equation}
with  $\alpha/2>\lambda>\beta/2$. The eigenvalues are:
$$e^{\frac{2\pi i k}{3}}\lambda\,\,,\,\,\,k=0,1,2\,.$$
In the limit $\lambda\rightarrow 0$ we recover the $Q^3=0$ case.  The solutions are complicated and presented in Appendix \ref{app:generalQ}. They are again plagued by singularities that appear unphysical and are of the ghost brane type\footnote{This for instance implies that the singularities violate both Gubser's and the Maldacena-Nunez criteria \cite{Gubser:2000nd, Maldacena:2000mw}}. This is due to the geometric functions (sines and cosines) that appear. Their presence is a simple consequence of the complex eigenvalues of the $Q$-matrix and cannot be avoided. The $\lambda$-deformation to the nilpotent $Q$-matrix does not help cure the singularities associated to ghost branes, as it only introduces extra singularities.\\

We conclude that the problem of finding well behaved D2-D4 brane bound state solutions is still an open one. We foresee two possibilities. First, that there may be no solution in supergravity since this bound state is not consistent by itself. This could happen because of non-zero forces on the brane stacks which would imply that any physical solution would be time-dependent. Second, that a static solution does exist but one would have to move beyond the simple Ansatz we made. This can happen if there are non-trivial effects taking place. For instance, one can imagine that the D2 brane wants to polarise into a spherical D4 brane under the influence of the $F_4$-fluxes sourced by the D4 stacks. We leave an investigation of such potential effects for the future.

\section{The \texorpdfstring{D$(-1)$-D7}{D-1D7} bound state} \label{sec:D-1-D7}
We consider a simplified version of the Ansatz from \cite{Aguilar-Gutierrez:2022kvk}\footnote{The aim of \cite{Aguilar-Gutierrez:2022kvk} was to find a near-horizon solution. Here, we include $y$-dependence to find a full solution.} which could capture a D$(-1)$-D7 bound state with the D7 branes wrapped over an 8-torus and the D$(-1)$ branes smeared over the same torus:
\begin{align}
    \dd s^2 &= M_x(y)^2 \dd x^2 + M^2_y(y)\dd y^2 + M_1(y)^2\sum_{i=1}^4 (\dd \theta^i)^2 + M_5(y)^2\sum_{i=5}^8 (\dd \theta^i)^2\,, \label{eq:ansatz_D-1D7_metric}\\
    F_5 &= (1-i\hodge)\mathcal{F}, \quad \mathcal{F}=\dd \theta^{1234} \wedge (\gamma(y)\dd x + i\delta(y)\dd y) \label{eq:ansatz_D-1D7_F5}\\
    F_1 &= \alpha(y)\dd x + i \beta(y) \dd y \label{eq:ansatz_D-1D7_F1}\,.
\end{align}
The metric is written in Einstein frame\footnote{Whereas \cite{Aguilar-Gutierrez:2022kvk} used string frame and the relation between the variables $L$ of \cite{Aguilar-Gutierrez:2022kvk} is $M^2=e^{-\phi/2}L^2$).}. The coordinates $\theta$ parametrize the $\mathbb{T}^8$, the $x$ direction is also considered to be a circle with $2\pi$ periodicity, while $y$ is a noncompact direction corresponding to the would-be ``AdS$_1$'' factor.\footnote{Supergravity equations are blind to such topological statements.} In what follows we will regard $y$ as Euclidean time. We assume the dilaton only depends on the $y$-coordinate and we hope to find a constant dilaton solution corresponding to the near horizon of the bound state. \\

Without the $F_5$ fluxes, we do not expect to find physical solutions with both D$(-1)$ and D7 charges as explained in \cite{Aguilar-Gutierrez:2022kvk}; for bound states where the the number of mixed directions is eight there are (T-duals of) Hanany-Witten effects taking place \cite{Hanany:1996ie, Danielsson:1997wq}. A good example is the supergravity solution for a  D0$-$D8 bound state \cite{Massar:1999sb, Imamura:2001cr, Janssen:1999sa,Bergshoeff:2003sy}.
The natural Ansatz for a D0$-$D8 bound state, ignoring such effects would be a three-block Ansatz with the harmonic functions in the right places, but it indeed does not solve the equations of motion. A dual version of the Hanany-Witten effect suggests that a fundamental string (denoted F1) is created, stretching between the $D0$ and the $D8$. Once this F1 is included as in the table below, 
\begin{equation}
    \begin{aligned}
D0 \quad  & \times - - - - - - - - -  \\
F1 \quad  & \times - - - - - - - - \times \\
D8\quad &  \times \times \times \times \times \times 
\times \times \times - \\
\end{aligned}
\end{equation}
solutions to the EOMs can be found \cite{Massar:1999sb, Imamura:2001cr, Janssen:1999sa,Bergshoeff:2003sy}. For the D$(-1)$-D7 bound state one expects that worldvolume fluxes on the D7 brane are needed \cite{Billo:2021xzh}. They would act like (dissolved) D3 branes \cite{Aguilar-Gutierrez:2022kvk}, explaining the $F_5$ fluxes in our Ansatz \eqref{eq:ansatz_D-1D7_F1}.\\

Finally, a word on the appearance of the imaginary unit $i$ in our Ansatz. Since Euclidean gravity by itself is already a Wick rotation of Lorentzian gravity obtained by making time imaginary there is a tendency to think that general complex field configurations are allowed as saddle points in the path integral. However, since one cannot just double all degrees of freedom, there are subtleties. We follow the logic of reality conditions in Euclidean supergravity, which relies on taking real sections such that the action remains bounded from below (possibly up to the conformal factor problem if it appears). A concise summary of this can be found in section 2 of \cite{Aguilar-Gutierrez:2022kvk}. The rule of thumb is that magnetic charges are real (and consequently electric charges are imaginary), as is explicit for instance in the expression for $F_1$ in our Ansatz \eqref{eq:ansatz_D-1D7_F1}. A D$(-1)$-brane is electrically charged under $F_1$ and magnetic under $F_9$ and vice versa for D7 branes. Hence we think of $\alpha$ as describing D7 charge and $\beta$ D$(-1)$ charge (recall that $y$ is Euclidean time). Insisting on real magnetic D$(-1)$-D7 charges then implies $\alpha$ and $\beta$ are real.  A similar logic applies to the $F_5$-flux \eqref{eq:ansatz_D-1D7_F5} where $\gamma$ and $\delta$ are taken real in order to represent real D3 charges. Note, however, that reference \cite{Aguilar-Gutierrez:2022kvk} took opposite reality conditions for $\gamma$ and $\delta$ and this is where we differ with \cite{Aguilar-Gutierrez:2022kvk} in what follows.\footnote{Note that we decided to also use a different notation for $\delta$ here, which is related to $i\delta$ in \cite{Aguilar-Gutierrez:2022kvk}. This notational difference reflects the different reality conditions. We will therefore find different solutions for the D$(-1)$-D7 bound state, and as we explain below, we will also differ on the supersymmetry variations.}

\subsection{An effective action for the \texorpdfstring{D$(-1)$-D7}{D-1D7} system}
In what follows we construct a 1d effective action for the variables in the Ansatz, which will be written in the form:
\begin{equation}\label{EFFACT}
S  =\int d y \Bigl( -\frac{1}{2}G_{ij}\dot{\Phi}^i \dot{\Phi}^j - V(\Phi) \Bigr)   \,, 
\end{equation} 
where $\Phi^i$ are suitable field redefinitions of variables in the Ansatz.

Let us start with the kinetic terms. For the $M$-variables, we can obtain them from dimensionally reducing the Einstein-Hilbert term. After dropping boundary terms we find
\begin{equation}
S_{kin}=\int dy M_1^4M_5^4\frac{M_x}{M_y}\Bigl(12 \frac{M_1'^2}{M_1^2} + 12 \frac{M_5'^2}{M_5^2} + 32 \frac{M_1'M_5'}{M_1M_5} + 8 \frac{M_x'M_5'}{M_xM_5} +  8 \frac{M_x'M_1'}{M_xM_1} -\frac{1}{2}\phi'^2 \Bigr)\,,  
\end{equation}
and we added the dilaton kinetic term. We can find variables in which the kinetic term is diagonal and flat (and of indefinite signature) but it turns out that other variables, in which the kinetic term is not diagonal, are more useful. First, we choose the gauge
\begin{equation}\label{My}
M_y = M_xM_1^4M_5^4\,,    
\end{equation}
and then introduce the variables $u, v, \chi_1$ as follows
\begin{align}
    M_x^2 &= e^{\frac{1}{2}\frac{1}{\omega}v}e^{-\frac{7}{4}\omega u} \,,\label{eq:D(-1)D7_uplift_formula_Mx}\\
    M_y^2 &= e^{\frac{1}{2}\frac{1}{\omega}v}e^{\frac{9}{4}\omega u} \,,\label{eq:D(-1)D7_uplift_formula_My}\\
    M_1^2 &= e^{\frac{1}{2}\chi_1}e^{\frac{1}{2}\omega u} \,,\label{eq:D(-1)D7_uplift_formula_M1}\\
    M_5^2 &= e^{-\frac{1}{2}\chi_1}e^{\frac{1}{2}\omega u} \,,\label{eq:D(-1)D7_uplift_formula_M5}
\end{align}
where $\omega = (57/4)^{-1/4}$.
The kinetic term of our effective action \eqref{EFFACT}  
is now given by
\begin{equation}
 G_{ij} =
 \begin{pmatrix} 
 1 & 0 & 0 & 0 \\
 0 & 0 & -1 & 0 \\
 0 & -1 & 0 & 0\\
 0 & 0 & 0 & 1 
 \end{pmatrix}
 = G^{ij}    \,,
\end{equation}
with $\Phi^i= (\chi_1,u,v,\phi)$. 

We now turn to computing $V$, the potential for the 1d system \eqref{EFFACT} which originates from reducing the fluxes, so $V=V_{F1}+ V_{F_5}+ V_{F_9}$. From reducing the magnetic contribution of $F_1$ we get
\begin{equation}
        V_{F_1} = \frac{1}{2}M_1^4 M_5^4 M^{-1}_x M_y e^{2\phi}\alpha^2\,.
\end{equation}
To reduce the magnetic piece of $F_9$ we need to be careful. Note that
\begin{equation}
 F_9 = \beta e^{2\phi}M_x M_y^{-1}M_1^4M_5^4 \dd x \wedge \dd\theta_1\ldots \dd\theta_8\,.   
\end{equation}
Given the $F_1$ equation of motion, we do not keep $\beta$ fixed but the whole coefficient
\begin{equation}
\tilde{\beta} = \beta e^{2\phi} M_x M_y^{-1}M_1^4M_5^4 \label{eq:D-1D7_beta_betatilde}\,,     
\end{equation}
which leads to:
\begin{equation}
    V_{F_9} = \frac{1}{2}M_1^{-4} M_5^{-4} M^{-1}_x M_y e^{-2\phi} \tilde{\beta}^2\,. 
\end{equation}
A similar procedure for $F_5$
\begin{equation}
F_5 = (1-i\star) \left(\gamma \dd x \wedge \dd \theta^{1234} -\delta \frac{M_x}{M_y}\left(\frac{M_5}{M_1}\right)^4 \dd x \wedge \dd \theta^{5678}\right) \,,    \end{equation}
implies we keep $\gamma$ fixed and 
\begin{equation}
\tilde{\delta} = \delta \frac{M_x}{M_y}\left(\frac{M_5}{M_1}\right)^4\,\label{eq:D-1D7_delta_deltatilde}.
\end{equation}
Such that we find:
\begin{equation}
V_{F_5} =  \frac{1}{2}M_x^{-1} M_y\left( \frac{M_5^4}{M_1^4} \gamma^2 + \frac{M_1^4}{M_5^4}\tilde{\delta}^2 \right)\,.   
\end{equation}
The total potential then is
\begin{equation}\label{V(M)}
V = \frac{1}{2}\frac{M_y}{M_x}\left(M_1^4 M_5^4e^{2\phi}\alpha^2 + M_1^{-4} M_5^{-4} e^{-2\phi} \tilde{\beta}^2+  \frac{M_5^4}{M_1^4} \gamma^2 + \frac{M_1^4}{M_5^4}\tilde{\delta}^2 \right)  \,.  
\end{equation}
This potential is positive definite for the choices of reality conditions we insist on. The equations of motion from the effective potential reduce to the 10d equations of motion on the conditions that
\begin{align}\label{eq:D(-1)D7_flux_constraint}
    \alpha\tilde{\beta}+\gamma\Tilde{\delta} = 0\,.
\end{align}
In other words, only 3 flux quanta in the potential are free to chose. This constraint is enforced by the 10d reversed Einstein equation for $R_{xy}$ and needs to be imposed in the definition of the 1d effective potential. Its physical interpretation is exactly the dual to the Hanany-Witten effect we alluded to above. We notice that whenever we have both D$(-1)$ and D$7$ charges (related to $\alpha$ and $\beta$) we need to introduce D3 charges (related to $\gamma, \delta$).

\subsection{Supersymmetry and superpotential}
We would like to build an effective (super)potential $W$ for the effective action found above in terms of the  fields $\Phi^i= (\chi_1,u,v,\phi)$.  If the potential $V$ can be expressed through a superpotential $W$ as
\begin{align}
    V = \frac{1}{2}G^{ij}\partial_i W \partial_j W\, , \label{eq:superpotential_potential_relation}
\end{align}
then the action reduces to a square up to boundary terms
\begin{align}
    S &= -\frac{1}{2}\int \dd y \ \left(\dot{\Phi}^i+G^{ij}\partial_i W\right)^2 + \int \dd y \ \frac{\dd W}{\dd y} \,.
\end{align}
The equations of motion of the effective action are then implied by the first order equations
\begin{align}
    \dot{\Phi}^i+G^{ij}\partial_i W &= 0 \,.\label{eq:superpotential_def_scalars}
\end{align}
To find first-order equations (and thus a $W$-function) we consider the supersymmetry variation of the gravitino and the dilatino \cite{Aguilar-Gutierrez:2022kvk}. Since SUSY variations are usually presented in string frame, we present the BPS equations in terms of the string frame variables $L$ which are related to the $M$-variables as: $M^2=e^{-\phi/2}L^2$. The first-order equations we find are:
\begin{align}
    \phi' &= \eta_d e^{-\phi}\left(\alpha L_1^4L_5^4 -\eta_d \tilde{\beta}\right)\,,\\
    \eta_d L_x'  &= -\frac{1}{4}L_x e^{-\phi}\left[\left(\alpha L_1^4L_5^4  + \eta_d \tilde{\beta} \right)+i^z\eta_p\left(\gamma L_5^4 +\eta_d\tilde{\delta}L_1^4\right)
    \right]\,,\\
    \eta_d L_1' &= -\frac{1}{4}L_1 e^{-\phi}\left[-\left(\alpha L_1^4L_5^4- \eta_d \tilde{\beta} \right)
    + i^z\eta_p\left(\gamma L_5^4 -\eta_d\tilde{\delta} L_1^4 \right)\right]\,,\\
    \eta_d L_5' &= -\frac{1}{4}L_5 e^{-\phi}\left[-\left(\alpha L_1^4L_5^4 - \eta_d\tilde{\beta}\right) 
    - i^z\eta_p\left(\gamma L_5^4 -\eta_d\tilde{\delta}L_1^4\right)\right] \,.
\end{align}
The derivation can be found in Appendix \ref{app:details_D-1D7}.
The above is expressed with the same gauge choice ($M_y=M_xM_1^4M_5^4$) as for the effective action. The parameter $z \in \{0,1\}$, introduced in \cite{Aguilar-Gutierrez:2022kvk}, represents the supposed ambiguity in defining the supersymmetry variations in Euclidean signature. The variables $\eta_{p,d}$ verify $\eta^2=1$ and are simply sign choices in the definition of the projectors on the SUSY parameter:
\begin{align}
    (\Gamma_{\Bar{x}} + i \eta_d \Gamma_{\Bar{y}}) \epsilon &= 0\,,\\
    \Gamma_{\bar{1}\bar{2}\bar{3}\bar{4}}\epsilon &= \eta_p \epsilon\,.
\end{align}
Our goal is to fix the ambiguity in the supersymmetry variations by demanding that the first-order equations square to the equations of motion. This will fix $z$ to $z=0$ \emph{different from the choice made in \cite{Aguilar-Gutierrez:2022kvk}}.
In terms of the canonically normalized fields, the first-order equations (imposed by supersymmetry) can be derived from the following superpotential:
\begin{equation}\label{Wfunction}
W = -\eta_d \alpha e^{\phi + 2\omega u} - \tilde{\beta}e^{-\phi} - \eta_p \left(\eta_d\gamma e^{-\chi_1}+\tilde{\delta} e^{\chi_1} \right) e^{\omega u}\, ,    
\end{equation}
via equation \eqref{eq:superpotential_def_scalars}. This $W$-function should reproduce our effective potential, according to equation \eqref{eq:superpotential_potential_relation}. One finds this only happens when $z=0$ and when the constraint on flux quanta
\eqref{eq:D(-1)D7_flux_constraint} is satisfied. Given that the effective action is a consistent truncation of the 10d theory, this shows that the first order Killing spinor equations square to the second order equations of motion provided that $z=0$.

It turns out that the $W$-function contains all the relevant physics. First of all, since the variables only depend on $y$ we expect the sources to be smeared over the $x$-direction and they become effectively co-dimension one. To interpret the flow solutions consistently as branes solutions, i.e. to change fluxes for branes, we need to use Heaviside functions for the charges $\alpha,\tilde{\beta}, \gamma, \tilde{\delta}$.  We will take the convention where all flux will be traded by branes, meaning that all charges should vanish for $y<y_0$ for some $y_0$. Then, just like the D8 brane in massive IIA, the brane tension is the value of $W$ at the jump \cite{Bergshoeff:1996ui}. Then $W$ becomes the sum of all brane tensions (in Einstein frame). To see this note that $e^{2\omega u}$ is the volume of the $\mathbb{T}^8$ and the volumes of the separate $\mathbb{T}^4$'s are $e^{\pm\chi_1+\omega u}$. We also have that, in Einstein frame, D$p$ tensions scale as  $e^{(p-3)\phi/4}$ times their charges. Our superpotential is then exactly the sum of the tensions of a (smeared) D$(-1)$ stack, D7 stack wrapping $\mathbb{T}^8$ a D3 stack wrapping one $\mathbb{T}^4$ and a D3 stack wrapping the other $\mathbb{T}^4$:
\begin{equation}
    \begin{aligned}
D(-1) \quad  & - - - - - - - - - -  \\
D7 \quad  & - - \times \times \times \times \times \times 
\times \times  \\
D3\quad &  - - \times \times \times \times - - - - \\
D3\quad &  \underbrace{-\,\, -}_{y, x} \underbrace{ - - - \,- }_{\theta_1,\ldots,\theta_4} \underbrace{\,\times \times \,\times\,\times}_{\theta_5,\ldots,\theta_8} \\
\end{aligned}
\end{equation}
Before we solve the flow equations, which will further confirm the above physical picture, we can already discuss a crucial issue regarding the signs of the brane actions. For that consider equation \eqref{eq:D(-1)D7_flux_constraint} and multiply it with $\eta_d$: 
\begin{equation}
(\eta_d\alpha)\tilde{\beta}+(\eta_d\gamma)\Tilde{\delta} = 0\,.    
\end{equation}
Note that the $W$-function is determined up to a minus sign, which reflects the choice of $y$-axis orientation. When we want the D$(-1)$ and D7 sources to have positive tension we chose $\eta_d\alpha$ and $\beta$ to be negative. But then we notice that the product of the two stacks of D3 tensions ($(\eta_d\gamma)\Tilde{\delta}$) needs to be negative, regardless of the sign of $\eta_p$. Hence we conclude that one of the D3 stacks must have negative tension. They are so-called ghost branes. But in Euclidean signature this does not have to pose a problem. Even more, it is needed to create an object with vanishing total action (tension). The latter has been conjectured to be a hallmark of a conformal matrix theory in zero dimensions \cite{Aguilar-Gutierrez:2022kvk}.\footnote{We are grateful to Nikolay Bobev for that suggestion.}

To verify the above physical picture we truncate some charges in order to recognize the separate brane stacks as solutions to the flow equations. Starting from the flow equations \eqref{eq:D(-1)D7_flow_chi1}-\eqref{eq:D(-1)D7_flow_phi}, one expects to find the D$(-1)$ brane solution if $\alpha=0$, and the D7 brane solution if $\tilde{\beta}=0$. One furthermore has to consider the presence of D3 branes since $F_5$ need not vanish. Indeed, the flux quanta condition \eqref{eq:D(-1)D7_flux_constraint} tells us that if the product $\alpha\tilde{\beta}=0$, then $\gamma\tilde{\delta}=0$. If we choose both $\gamma$ and $\tilde{\delta}$ to vanish, then we expect to find the D$(-1)$ or the D7 brane solution. However, if either of these $F_5$ charges is non-zero, we expect to find bound states of the form D$(-1)-$D3 or D3$-$D7. We now verify this and solve the flow equations (with $z=0$) for the subsets of the charges and after that we solve for the general 4-charge solution. When written out, the flow equations are
\begin{align}
    \dot{\chi_1} &= -\eta_p \left(\eta_d\gamma e^{-\chi_1}-\tilde{\delta} e^{\chi_1} \right) e^{\omega u} \label{eq:D(-1)D7_flow_chi1}\, ,\\
    \dot{u} &= 0 \label{eq:D(-1)D7_flow_u}\, ,\\
    \dot{v} &= -2\eta_d \omega \alpha 
    e^{\phi
        + 2\omega u}
    - \eta_p \omega \left(\eta_d\gamma e^{-\chi}+\tilde{\delta} e^{\chi_1} \right) e^{\omega u} \label{eq:D(-1)D7_flow_v}\, ,\\
    \dot{\phi} &= \eta_d \alpha 
    e^{\phi
        + 2\omega u}
    - \tilde{\beta}e^{-\phi} \label{eq:D(-1)D7_flow_phi} \, .
\end{align}

\paragraph{D$(-1)$ and D7 branes, separately} \ \\
We start with setting $\alpha=\gamma=\tilde{\delta}=0$. The flow equations are solved trivially as $\chi_1$, $u$ and $v$ are just integration constants, and the dilaton profile is
\begin{align}
    e^{\phi} &= C_{\phi} - \tilde{\beta}y\, ,
\end{align}
where again $C_{\phi}$ is an integration constant. The uplift of this solution is
\begin{align}
    \dd s_s^2 &= e^{\phi/2}\left(M_x^2 \dd x^2 + M_y^2 \dd y^2 + M_1^2 \dd \theta_{1234}^2 + M_5^2 \dd \theta_{5678}^2\right)\,,\\
    F_1 &= i\tilde{\beta} e^{-2\phi} \dd y\,,
\end{align}
where the metric is written in string frame. Now $M_x, M_1$ and $M_5$ are integration constants, and $M_y$ is fixed by our earlier gauge choice to be $M_y=M_xM_1^4M_5^4$. We can set them all to 1. Then we find the usual string frame solution for a stack of D$(-1)$ branes in terms of the harmonic function $H_{-1}(y) = 1 - \tilde{\beta} y$. Yet, the harmonic corresponds to that of a codimension one object. This implies that the D-instantons are not just smeared along the $\mathbb{T}^8$ (the worldvolume of the D7 branes), but also along $x$. We do not know why this happens since our Ansatz is general enough for a localised solution where $y$ could be a radial coordinate and $x$ an angle on a 2d plane. 

Let us now set $\tilde{\beta}=\gamma=\tilde{\delta}=0$ and solve the flow equations to obtain constant $\chi_1$ and $u$ functions, as well as
\begin{align}
    e^{\frac{1}{\omega}v} &= A_v e^{-2\phi}\\
    e^{-\phi} &= C_{\phi}-\eta_d \alpha e^{2\omega u}y
\end{align}
where $A_v$ and $C_{\phi}$ are integration constants. 
As before, we may choose integration constants such that $A_v=C_{\phi}=1$, $\chi_1=u=0$, which yields the usual (Euclidean) D7 expression smeared over the $x$ coordinate:
\begin{align}
    \dd s_s^2 &= H_7^{1/2}\left(\dd x^2 +  \dd y^2\right) + H_7^{-1/2}\dd \mathbb{T}_8^2, \quad H_7 = 1-\eta_d \alpha y\, ,\\
    e^{\phi} &= H_7^{-1}\qquad F_1 = \alpha \dd x\, .
\end{align}

\paragraph{D$(-1)-$D3 and D3$-$D7 bound states} \ \\
Let us now consider the case where we do not set both $\gamma$ and $\tilde{\delta}$ to 0, but only one of them. 
We begin by considering the D$(-1)-$D3 case ($\alpha=0$).  Solving the flow equations with $\gamma=0$ leads to (after fixing integration constants as before):
    \begin{align}
        \dd s_s^2 &= H_{-1}^{1/2} H_3^{1/2}(\dd x^2 + \dd y^2 + \dd\theta_{5678}^2)+H_{-1}^{1/2} H_3^{-1/2}\dd\theta_{1234}^2\\
        H_{-1} &= 1-\tilde{\beta}y\, ,\quad H_3 = 1 - \eta_p (\tilde{\delta}) y \, ,\\
        F_5 &= \tilde{\delta} \left(\dd x\wedge \dd \theta^{5678} + i H_3^{-2} \dd y\wedge \dd \theta^{1234}\right)\\
        F_1 &= i\tilde{\beta}H_{-1}^{-2} \dd y\, \\
        e^{\phi} &= H_{-1}
    \end{align}
    This matches our expectation of the bound state of a D-instanton and a D3 brane in the $\theta^{1234}$ directions.
When instead $\tilde{\delta}=0$, we find a similar solution with a D3 stack along the $\theta^{5678}$ directions, described by the harmonic $H_3 = 1 - \eta_p \eta_d \gamma y $ and $F_5$ profile given by
\begin{equation}
    F_5 = \gamma \left(\dd x\wedge \dd \theta^{1234}
        + i H_3^{-2} \dd y\wedge \dd \theta^{5678}\right)
\end{equation}
A similar computation can be done for the D3$-$D7 bound state and leads to the usual harmonic superposition rules.  

\subsection{Solving the full flow equations}
We demonstrated that the D$(-1)$ and D7 stacks separately, as well as the D$(-1)-$D3 and the D3$-$D7 bound states can be obtained from the flow equations. The corresponding harmonic functions were:
\begin{align}
    D(-1) :\ &H_{-1} = 1-\tilde{\beta} y\, ,\\
    D7 :\ &H_{7} = 1-\eta_d \alpha y\, ,\\
    D3 :\ &H_3^{(1234)} = 1 - \eta_p \tilde{\delta} y\, ,\\
    &H_3^{(5678)} = 1 - \eta_p \eta_d \gamma y\, .
\end{align}
The D$(-1)$ and D7 should have positive tension, so we restrict to $\tilde{\beta}<0$, $\eta_d \alpha < 0$.  Now that we have fixed the signs, equations \eqref{eq:D(-1)D7_flow_chi1} and \eqref{eq:D(-1)D7_flow_phi} can be solved easily for the general 4-charge solution. \\

The flow equations imply that $u(y)$ is constant, so we set it to $u_0$. This represents the volume of the 8-torus, since $\operatorname{vol}(\mathbb{T}_8)=e^{2\omega u_0}$. Instead of using $u_0$, we now use $\mathcal{V} \equiv e^{2\omega u_0}$ since it has a physical meaning. There are two branches of solutions for each equation, which will be selected upon choosing appropriate boundary conditions. We find:
\begin{align}
        e^{\phi} &= \left\{\begin{aligned}
            \mathcal{V}^{-1/2}\sqrt{\frac{\tilde{\beta}}{\eta_d\alpha}}\tanh\left(\mathcal{V}^{1/2}\sqrt{\eta_d\alpha\tilde{\beta}}(y+C_{\phi})\right)\\
            \mathcal{V}^{-1/2}\sqrt{\frac{\tilde{\beta}}{\eta_d\alpha }}\coth\left(\mathcal{V}^{1/2}\sqrt{\eta_d\alpha\tilde{\beta}}(y+C_{\phi})\right)\\
        \end{aligned} \right. \, ,
        \\
        e^{\chi_1} &= \left\{\begin{aligned}
            -\sqrt{\frac{-\eta_d \gamma}{\tilde{\delta}}}\cot\left(\mathcal{V}^{1/2}\sqrt{-\eta_d \gamma\tilde{\delta}}(y+C_{\chi})\right)\\
             \sqrt{\frac{-\eta_d \gamma}{\tilde{\delta}}}\tan\left(\mathcal{V}^{1/2}\sqrt{-\eta_d \gamma\tilde{\delta}}(y+C_{\chi})\right)\\
     \end{aligned} \right. \, ,
\end{align}
where $C_{\phi}$ and $C_{\chi}$ are integration constants. With these in hand, we find a solution for $v$:
\begin{align}
    e^{\frac{1}{\omega} v} &= A_v \left|\sin\left(2\mathcal{V}^{1/2}\sqrt{-\eta_d \gamma\tilde{\delta}}(y+C_{\chi})\right)\right| 
    \left\{\begin{aligned}
            \cosh^2\left(\mathcal{V}^{1/2} \sqrt{\eta_d\alpha\tilde{\beta}}(y+C_{\phi})\right)\\
            \sinh^2\left(\mathcal{V}^{1/2} \sqrt{\eta_d\alpha\tilde{\beta}}(y+C_{\phi})\right)\\
        \end{aligned} \right.\, ,
\end{align}
where now the two branches of $v$ solutions are correlated with the two branches of $\phi$ solutions. Note that both branches of $\chi_1$ solutions lead to the same $v$. We will not attempt to discuss which of the two branches are physical. We prefer to wait until localised solutions are constructed.\\

The uplift to 10 dimensions can be done easily using \eqref{eq:D(-1)D7_uplift_formula_Mx}-\eqref{eq:D(-1)D7_uplift_formula_M5}. Concerning singularities, a few words are in order. We have not been careful in stating what the local jumps in the fluxes are. This is something that can be chosen and will determine the effective charge/tension (action) of the sources. As with all co-dimension one sources singularities at finite distances are expected, and indeed observed here,  because the backreaction of such objects does not deplete but rather grows away from the source. Whether globally consistent solutions can be found by inserting orientifolds goes beyond the scope of this work. Our main goal was demonstrating that bound state solutions could be found that solve the SUSY variations and equations of motion on the condition that D3 charges are included. Our ultimate goal is understanding whether a smooth near-horizon solution is possible. One would expect such a solution to have constant dilaton. We will turn to constant dilaton solutions next, but we should keep in mind that smooth near horizon solutions are not to be expected within this approach because of the smearing along the $x$-direction. For instance, the near horizon limit of smeared D3 branes does not lead to $AdS_5\times S^5$. 

Instead we consider it worthy to pursue the search of solutions of a similar Ansatz but where variables depend both on $x$ and $y$. This would make the branes involved backreact as co-dimension two sources, and in analogy with D7 solutions \cite{Gibbons_1996}, one could expect the variables to have holomorphic dependence only, ie, in a complex parametrisation of the two-dimensional space spanned by $x,y$ we expect the solutions to only depend on $x+iy$ and we hope to report on this in the future. 

\subsection{\texorpdfstring{D$(-1)$-D7}{D-1D7} near horizon?}
As we explained, the near horizon geometry of D3, D1-D5 and D0-D6 stacks are vacuum solutions by themselves of the form $AdS_{p+2}\times S_{p+2}\times \mathbb{T}^{6-2p}$ for $p=3,1,0$. When $p=2$ we found no such solution.\\

In reference \cite{Aguilar-Gutierrez:2022kvk} the question was raised whether the D$(-1)$-D7 bound state would similarly lead to a vacuum solution that could be considered $AdS_{1}\times S_{1}\times \mathbb{T}^{8}$. But since there is no curvature in 1d, the $AdS_1$ part was just interpreted as a flat line ($\mathbb{R}_1$). Indeed, a solution of that form was found with imaginary self-dual $F_1$ flux along the $AdS_1\times S^1$ part. To reproduce this solution using our effective potential method we simply verify whether $\partial_i V$ can be made to vanish such that all fields are constant. For this purpose it is convenient to use \eqref{V(M)}. It is not difficult to verify that this is possible for the reality conditions taken in \cite{Aguilar-Gutierrez:2022kvk} which have imaginary $\gamma$ and $\delta$.\\

It was argued in \cite{Aguilar-Gutierrez:2022kvk} that this solution preserves supersymmetry and hence should correspond to the near horizon of the would-be D$(-1)$-D7 bound state (with D3 branes dissolved in flux along the $\mathbb{T}^8$). In the previous subsections instead we have found the full D$(-1)$-D7 bound state and we have shown that the conjecture of \cite{Aguilar-Gutierrez:2022kvk} that the vacuum solution $AdS_{1}\times S_{1}\times \mathbb{T}^{8}$ is the near horizon seems false when we restrict to the SUSY variations with $z=0$. The reason for the difference between the results presented here and in \cite{Aguilar-Gutierrez:2022kvk} boil down to the difference in SUSY variations ($z=1$ vs $z=0$) \emph{and} the reality conditions for $F_5$. We have explained earlier that the choice $z=1$ seems inconsistent as the first-order equations of bound state solutions are not consistent with the second-order equations of motion when $z=1$.  We conclude that it is likely that the vacuum solution found in \cite{Aguilar-Gutierrez:2022kvk} is not the physical  holographic background for the D$(-1)$-D7 matrix theory of \cite{Billo:2021xzh}.

Although we do not expect a well behaved near horizon due to the smearing along $x$ we can still try to search for constant dilaton solutions with our current choice of reality conditions for the fluxes and with SUSY defined by $z=0$ instead of $z=1$. Such solutions could serve as a hallmark of a would-be near horizon. Assuming $\phi(y)=\phi_0$, in the flow equations, we find the following general solution:
\begin{align}
    e^{\chi_1} &= \left\{\begin{aligned}
            -\sqrt{\frac{-\eta_d\gamma}{\tilde{\delta}}}\cot\left(\mathcal{V}^{1/2} \sqrt{-\eta_d\gamma\tilde{\delta}}(y-y_0)\right)\\
            \sqrt{\frac{-\eta_d\gamma}{\tilde{\delta}}}\tan\left(\mathcal{V}^{1/2}\sqrt{-\eta_d\gamma\tilde{\delta}}(y-y_0)\right)\\
     \end{aligned} \right. \, ,\\
    e^{\frac{1}{\omega}v} &= A_v e^{2 \mathcal{V}^{1/2}\sqrt{\eta_d\alpha\tilde{\beta}} y} \left|\sin\left(2\mathcal{V}^{1/2}\sqrt{-\eta_d\gamma\tilde{\delta}}(y-y_0)\right)\right|\, ,
\end{align}
where $A_v$ and $y_0$ are integration constants and the $\alpha$, $\tilde{\beta}$ charges are related to the dilaton value by
\begin{align}
    \eta_d \alpha e^{2\phi_0} \mathcal{V} = \tilde{\beta}
\end{align}
Note that this solution is the same as the one presented earlier if the dilaton integration constant $C_{\phi}$ is sent to $\pm \infty$. The 10d metric (in Einstein frame) of the constant dilaton solution is
\begin{align}
    \dd s^2 &= A_v^{1/2}e^{\sqrt{\eta_d\alpha\tilde{\beta}} y} \left|\sin\left(2\sqrt{-\eta_d\gamma\tilde{\delta}}(y-y_0)\right)\right|^{1/2}\left( \dd x^2 + \dd y^2\right)\nonumber \\
    &\hspace{1cm} + \left(\frac{-\eta_d\gamma}{\tilde{\delta}}\right)^{1/4}\left(\tan\left(\sqrt{-\eta_d\gamma\tilde{\delta}}(y-y_0)\right)\right)^{1/2}\sum_{i=1}^4 (\dd \theta^i)^2\nonumber \\
    &\hspace{1cm}+ \left(\frac{-\eta_d\gamma}{\tilde{\delta}}\right)^{-1/4}\left(\tan\left(\sqrt{-\eta_d\gamma\tilde{\delta}}(y-y_0)\right)\right)^{-1/2}\sum_{i=5}^8 (\dd \theta^i)^2\,,
\end{align}
where we set $\mathcal{V}=1$ for simplicity, and picked the ``tan'' branch for $\chi_1$. 
The same remarks concerning singularities we made for the full solution with non-constant dilaton apply here as well. We will not further investigate the physics of the solutions any further as we believe that the main goal was to show a SUSY solution with constant dilaton exists. We expect the more physical constant dilaton solution to come out when we localise the branes along the $x$-direction, which we leave for future research.

\section{Discussion}\label{sec:discussion}
We argued that it is natural to contemplate bound state configurations of D-branes where a brane is put next to its magnetic dual such that the brane with the most worldvolume dimensions is wrapping circles along the directions perpendicular to the smaller brane. One then obtains a dyonic object in the lower dimension and the near horizon should have constant dilaton. This is trivial for D3 branes as they are self dual and the near horizon is the famous $AdS_5\times S^5$ background. For D1-D5 bound states this gives dyonic strings in six dimensions whose near horizon describes the ten-dimensional geometry of the form $AdS_3\times S^4 \times \mathbb{T}^4$. 

In this paper we extended this for all D$p$ branes with $p\leq 7$. Such configurations are only expected to be SUSY for odd $p$. By lifting the dyonic Kaluza-Klein black hole we constructed the D0-D6 bound state with $AdS_2\times S^2 \times \mathbb{T}^6$ horizon. The real novel solutions arise for $p=-1$ and $p=2$.

Concerning the D2$-$D4 bound state we showed that supergravity solutions carrying the right charges can be constructed using group theory methods pioneered in \cite{Gibbons_1996, Bergshoeff:2008be}. This method relies on reducing the brane system to an instanton solution in a lower-dimension that is described by null geodesics on a scalar target space. This allows one to find all of the sought-for supergravity solutions carrying D$p$ and D$(6-p)$ charges, excluding however the D$(-1)-$D$7$ case.  We constructed all the null geodesics for the D2$-$D4 system, but the solutions suffer from singularities whose fate is not clear to us. In any case, we do \emph{not} find an $AdS_4\times S^4\times  \mathbb{T}^2$ near horizon. 

Concerning our D$(-1)$-D7 bound state solutions, our main goals were correcting and improving on reference \cite{Aguilar-Gutierrez:2022kvk} by 
\begin{enumerate}
\item settling the issue of the SUSY variations in Euclidean IIB with both D$(-1)$ and D7 flux. We believe we did so by fixing the ambiguities raised in \cite{Aguilar-Gutierrez:2022kvk}. Our method relied on demanding that first-order equations obtained from a Killing spinor analysis solved the second-order equations of motion. 
\item demonstrating the existence of SUSY bound state solutions carrying both D$(-1)$ and D7 charges. We found this is possible on the conditions D3 charges are generated as well. This is consistent with the string theory picture of \cite{Billo:2021xzh} which predicts worldvolume fluxes on the D7 stack, which were observed to induce D3 charges in \cite{Aguilar-Gutierrez:2022kvk}.
\item investigating the near horizon geometry with the hope of uncovering an ``AdS$_1$''$\times S^1\times\mathbb{T}^8$ geometry which would be holographically dual to an extension of the IKKT matrix model \cite{Ishibashi:1996xs} constructed in \cite{Billo:2021xzh} by allowing interactions between D$(-1)$ and D7 branes. Unfortunately our Ansatz only allowed for D$(-1)$-D7 branes along a line and so we do not expect it to be rich enough to construct this near horizon geometry. On the other hand we did demonstrate the existence of a constant dilaton solution, which could be seen as evidence in favor of a holographic near horizon geometry. The precise definition of conformally invariant matrix models is left for future research \cite{progress} as well as the construction of supergravity solutions with fully localised D7 branes. 
\end{enumerate}
Both novel classes of solutions presented in this paper, the ones with D$(-1)$-D7 charges and the ones with D2-D4 charges, feature ghost-like branes. For the D2-D4 solutions, this most likely indicates we are missing physical ingredients that can cure the associated singularities. We suggested this could be the polarisation of the D2 constituents into spherical D4 branes, which are described by a more complicated Ansatz than the one we considered. For the D$(-1)$-D7 solutions the singularities are all of the same type, namely singularities associated to co-dimension one objects. Such singularities for instance correspond to discrete jumps in flux parameters and the discrete difference in the superpotential at the jump represents the on-shell action and thus ``Euclidean tension''. We noticed that it can vanish due to the unavoidable presence of a ghost-like Euclidean D3 brane stack. Following \cite{Aguilar-Gutierrez:2022kvk} we suggest this is required to get the wanted holographic dual to a conformal matrix theory in zero dimensions. We leave a deeper understanding of this all for future research. 

\subsection*{Acknowledgments}
T.V.R. would like to acknowledge Klaas Parmentier and Sergio Aguilar for earlier work on the D$(-1)$-D7 supergravity solution. We furthermore acknowledge useful and inspiring discussions with Joel Karlsson and Nikolay Bobev. S.R. is supported by the Research Foundation - Flanders (FWO) doctoral fellowship 11PAA24N, and also acknowledges support from the FWO Odysseus grant G0F9516N and the KU Leuven iBOF-21-084 grant. The research of T.V.R. is in part supported by
the Odysseus grant GCD-D5133-G0H9318N of FWO-Vlaanderen.
\newpage

\appendix
\section{Uplifting the dilatonic black hole}
\label{app:dilatonic_BH_uplift}

We consider the 4d action
\begin{equation}
    S= \int\sqrt{|g|}\left\{R-\tfrac{1}{2}(\partial s)^2 -\tfrac{1}{4}e^{\sqrt3 s}F^2\right\}\,. 
\end{equation}
The following solution describes the well-known dilatonic black hole, see eg \cite{Chemissany:2010zp} and refs therein:
\begin{align}
& ds^2 = -e^{2U}dt^2 + e^{-2U}\left(dr^2 + r^2 \dd\Omega_2^2\right)\,,\qquad e^{2U} = (BC)^{-1/2}\,,\\
& e^{s} = (B/C)^{\frac{\sqrt3}{2}} \,,\\
& F= Q_M \dd \Omega_2 + Q_E \frac{e^{2U-a\phi}}{r^2} \dd t\wedge \dd r \,.
\end{align}
where 
\begin{align}
&   B  =  1 - \beta r^{-1} + \frac{\alpha\beta^2}{2(\alpha-\beta)}r^{-2}\,,\\
&   C  =  1 + \alpha r^{-1} - \frac{\alpha^2\beta}{2(\alpha-\beta)}r^{-2}\,. 
\end{align}
The constants $\alpha$ and $\beta$ can be written in terms of mass and charges as
\begin{equation}
 M = \frac{\sqrt2}{4}(\alpha-\beta)\,, \quad Q_E =\sqrt{\frac{\beta^3}{\beta-\alpha}}\,,\quad
 Q_M =\sqrt{\frac{\alpha^3}{\alpha-\beta}}
\end{equation}

Consider IIA string theory and truncate down to the bosonic action with RR 2-form field strength. This gives
\begin{equation}
S = \int\sqrt{-g}\left(R - \tfrac{1}{2}(\partial\phi)^2 -\tfrac{1}{4}e^{3\phi/2}F_2^2 \right)\,.
\end{equation}
We now reduce this action on a 6-torus keeping only the volume modulus of the torus as a consistent truncation:\begin{equation}
ds^2_{10}=e^{2\alpha\varphi}ds^2_4 +e^{2\beta\varphi}ds^2_6\,.
\end{equation}
where 4d Einstein frame requires
$3\beta=-\alpha$, and canonical normalisation of $\varphi$ requires $\alpha^2 =3/16$. The reduced action in 4d then is
\begin{equation}
S = \int\sqrt{-g}\left(R - \tfrac{1}{2}(\partial\phi)^2 - \tfrac{1}{2}(\partial\varphi)^2- \frac{1}{4}e^{3\phi/2-2\alpha\varphi}F_2^2 \right)\,.
\end{equation}
Consider the following rotation in field space
\begin{align}
s = \frac{1}{\sqrt3}\left(\frac{3}{2}\phi -2\alpha\varphi\right)\,,\\
t = \frac{1}{\sqrt3}\left(2\alpha\phi +\frac{3}{2}\varphi\right)\,.
\end{align}
We then get
\begin{equation}
S = \int\sqrt{-g}\left(R - \tfrac{1}{2}(\partial s)^2 - \tfrac{1}{2}(\partial t)^2- \tfrac{1}{4}e^{\sqrt3 s }F_2^2 \right)\,.
\end{equation}
The scalar $t$ decouples and we can put it to a constant  and then we recovered the dilatonic black hole Lagrangian \eqref{Einsteindilatonmaxwell}. This allows us to lift dilatonic black holes.

\section{Different branches of nilpotent \texorpdfstring{$Q$}{Q} matrices}\label{app:Qcube}

Note then that given a $Q$ matrix, we can generate others by defining
\begin{align}
    Q' = \Lambda^{-1} Q \Lambda \, , \ \text{where} \ \Lambda \eta \Lambda^{T} = \eta\, .
\end{align}
This is reminiscent of the disconnected components of the Lorentz group. The analogy of the parity and time reversal transformations here are the matrices
\begin{align}
    P = \operatorname{diag}(+1,-1,+1)\\
    T = \operatorname{diag}(-1,+1,+1).
\end{align}
Therefore, we can define "disconnected" $Q$ matrices as
\begin{align}
    Q_P = P Q P\, , \ 
    Q_T = T Q T\, , \
    Q_{PT} = TP Q PT\, .
\end{align}
The expressions for the fields are slightly altered, leading to different relations between $\alpha, \beta$ in terms of $Q_E, Q_M$:
\begin{align}
    Q_M^{(0)} &= \frac{3}{2}\frac{\alpha^{3/2}}{\sqrt{\alpha-\beta}}, & Q_E^{(0)} &= \frac{3}{2} \frac{(-\beta)^{3/2}}{\sqrt{\alpha-\beta}}\,,\\
    Q_M^{(P)} &= \frac{3}{2}\frac{\alpha^{3/2}}{\sqrt{\alpha-\beta}}, & Q_E^{(P)} &= -\frac{3}{2} \frac{(-\beta)^{3/2}}{\sqrt{\alpha-\beta}}\,,\\
    Q_M^{(T)} &= -\frac{3}{2}\frac{\alpha^{3/2}}{\sqrt{\alpha-\beta}}, & Q_E^{(T)} &= -\frac{3}{2} \frac{(-\beta)^{3/2}}{\sqrt{\alpha-\beta}}\,,\\
    Q_M^{(PT)} &= -\frac{3}{2}\frac{\alpha^{3/2}}{\sqrt{\alpha-\beta}}, & Q_E^{(PT)} &= \frac{3}{2} \frac{(-\beta)^{3/2}}{\sqrt{\alpha-\beta}}\,.
\end{align}
Note that for the normal convention (labeled by $(0)$) and $T$, the signs of $Q_E$ and $Q_M$ have to be equal, whereas for $P$ and $PT$, they have to be opposite. The sign of $B$ is equal to the sign of $Q_M$ for the normal and the $P$ case, whereas it is opposite for the $T$ and $PT$ cases.
Uplifting the solutions in which one truncates one charge to read off what brane solution is obtained one gets the usual D2 and D4 expressions with the following choice of harmonic functions: 
\begin{align}
    H_2^{(0)} &= 1-\frac{Q_E^{(0)}}{3r^3} & H_4^{(0)} &= 1+\frac{Q_M^{(0)}}{3r^3}\,,\\
    H_2^{(P)} &= 1+\frac{Q_E^{(P)}}{3r^3} & H_4^{(P)} &= 1+\frac{Q_M^{(P)}}{3r^3}\,,\\
    H_2^{(T)} &= 1+\frac{Q_E^{(T)}}{3r^3} & H_4^{(T)} &= 1-\frac{Q_M^{(T)}}{3r^3}\,,\\
    H_2^{(PT)} &= 1-\frac{Q_E^{(PT)}}{3r^3} & H_4^{(PT)} &= 1-\frac{Q_M^{(PT)}}{3r^3}\,.
\end{align}
We assume the notation is self-explanatory. Now recall that for $(0,T)$, $Q_E$ and $Q_M$ have the same sign whereas for $(P,PT)$, $Q_E$ and $Q_M$ have opposite sign. This means that the idea of using disconnected $Q$'s to get positive tension for both branes does not work. 

\section{General solutions for the null geodesics}\label{app:generalQ}
The general solutions to the extremal orbit, given by the $Q$-matrix \eqref{matQ} is:
{\scriptsize \begin{align}
    e^{\frac{\varphi}{\sqrt{3}}}&=\frac{\left[e^{\frac{\lambda  h }{2}} \left(\left(2 \lambda  (\alpha +\beta )-\alpha  \beta +8 \lambda ^2\right) \cos \left(\frac{1}{2} \sqrt{3} \lambda  h \right)+\sqrt{3} (2 \lambda  (\alpha +\beta )+\alpha  \beta ) \sin \left(\frac{1}{2} \sqrt{3} \lambda  h \right)\right)+(\alpha -2 \lambda ) (\beta -2 \lambda ) e^{-\lambda  h }\right]^{\frac{1}{2}}}{2\sqrt{3} |\lambda|}\,,\nonumber\\
    e^{-\Phi+\frac{\varphi}{\sqrt{3}}}&=\frac{\left(\alpha ^2+2 \alpha  \lambda +4 \lambda ^2\right) (2 \lambda -\beta ) e^{h \lambda }+(\alpha -2 \lambda ) e^{-\frac{1}{2} h \lambda } \left(\sqrt{3} \left(\alpha  (\beta +2 \lambda )+4 \lambda ^2\right) \sin \left(\frac{1}{2} \sqrt{3} h \lambda \right)+(\alpha  (\beta -2 \lambda )+4 \lambda  (\beta +\lambda )) \cos \left(\frac{1}{2} \sqrt{3} h \lambda \right)\right)}{12 \lambda ^2 (\alpha -\beta )}\,,\nonumber\\
    \chi_M&=\frac{\sqrt{\frac{\alpha ^3-8 \lambda ^3}{\alpha -\beta }} \left(\beta +e^{\frac{3 h \lambda }{2}} \left(\sqrt{3} (\beta +2 \lambda ) \sin \left(\frac{1}{2} \sqrt{3} h \lambda \right)-(\beta -2 \lambda ) \cos \left(\frac{1}{2} \sqrt{3} h \lambda \right)\right)-2 \lambda \right)}{(\alpha -2 \lambda ) (2 \lambda -\beta )+e^{\frac{3 h \lambda }{2}} \left(\left(-2 \lambda  (\alpha +\beta )+\alpha  \beta -8 \lambda ^2\right) \cos \left(\frac{1}{2} \sqrt{3} h \lambda \right)-\sqrt{3} (2 \lambda  (\alpha +\beta )+\alpha  \beta ) \sin \left(\frac{1}{2} \sqrt{3} h \lambda \right)\right)}\,,\nonumber\\
  \chi_E&=  \frac{\sqrt{-\frac{\beta ^3-8 \lambda ^3}{\alpha -\beta }} \left(\alpha +e^{\frac{3 h \lambda }{2}} \left(\sqrt{3} (\alpha +2 \lambda ) \sin \left(\frac{1}{2} \sqrt{3} h \lambda \right)-(\alpha -2 \lambda ) \cos \left(\frac{1}{2} \sqrt{3} h \lambda \right)\right)-2 \lambda \right)}{(\alpha -2 \lambda ) (2 \lambda -\beta )+e^{\frac{3 h \lambda }{2}} \left(\left(-2 \lambda  (\alpha +\beta )+\alpha  \beta -8 \lambda ^2\right) \cos \left(\frac{1}{2} \sqrt{3} h \lambda \right)-\sqrt{3} (2 \lambda  (\alpha +\beta )+\alpha  \beta ) \sin \left(\frac{1}{2} \sqrt{3} h \lambda \right)\right)}\,,\nonumber\\
  b&=\frac{\sqrt{\left(\alpha ^3-8 \lambda ^3\right) \left(8 \lambda ^3-\beta ^3\right)} \left(-e^{\frac{3 h \lambda }{2}}+\sqrt{3} \sin \left(\frac{1}{2} \sqrt{3} h \lambda \right)+\cos \left(\frac{1}{2} \sqrt{3} h \lambda \right)\right)}{\left(\alpha ^2+2 \alpha  \lambda +4 \lambda ^2\right) (2 \lambda -\beta ) e^{\frac{3 h \lambda }{2}}+(\alpha -2 \lambda ) \left(\sqrt{3} \left(\alpha  (\beta +2 \lambda )+4 \lambda ^2\right) \sin \left(\frac{1}{2} \sqrt{3} h \lambda \right)+(\alpha  (\beta -2 \lambda )+4 \lambda  (\beta +\lambda )) \cos \left(\frac{1}{2} \sqrt{3} h \lambda \right)\right)}\,.
\end{align}}

These complicated expressions for the fields can be simplified for specific choices of $\lambda$, but we refrain from presenting this since one can verify that the solutions are again plagued by singularities that appear unphysical.\\

The general solution we displayed above is found in terms of the $M$ entries, through:
\begin{align}
    \varphi&=\frac{1}{2} \sqrt{3} \log \left(M_{2,2} M_{3,3}-M_{2,3}^2\right)\,,\nonumber\\
    \Phi&=\log \left(\frac{\sqrt{M_{2,2} M_{3,3}-M_{2,3}^2}}{M_{3,3}}\right)\nonumber\\
    b&= \frac{M_{2,3}}{M_{3,3}}\,\,,\,\,\,\chi_E=\frac{M_{1,3} M_{2,3}-M_{1,2} M_{3,3}}{M_{2,3}^2-M_{2,2} M_{3,3}}\,,\nonumber\\
    \chi_M&= \frac{M_{1,3} M_{2,2}-M_{1,2} M_{2,3}}{M_{2,2} M_{3,3}-M_{2,3}^2}
\end{align}

\section{EOMs and SUSY of the \texorpdfstring{D$(-1)$-D7}{D-1D7} ansatz}
\label{app:details_D-1D7}

We summarize below the equations of motion and Killing spinor equations for our Ansatz (\eqref{eq:ansatz_D-1D7_metric}-\eqref{eq:ansatz_D-1D7_F1}). Recall that the string frame metric components are denoted by $L$'s, with the defining relation $M^2=e^{-\phi/2}L^2$.

\paragraph{Equations of motion}\ \\

The EOM and Bianchi identities for the gauge fields lead to
\begin{align}
    \beta M_x M_y^{-1} M_1^4 M_5^4 e^{2\phi} = \tilde{\beta}, \quad \delta M_x M_y^{-1} M_5^4 M_1^{-4}= \tilde{\delta}, \quad \alpha(y)=\alpha, \quad \gamma(y)=\gamma\, ,
\end{align}
where $\alpha, \tilde{\beta}, \gamma$ and $\tilde{\delta}$ are constants. The dilaton equation of motion is
\begin{align}
    \frac{M_x}{M_y}\frac{1}{M_1^4M_5^4} \partial_{y}\left(\frac{M_x}{M_y}M_1^4M_5^4\phi'\right) &= \alpha^2e^{2\phi}-\frac{1}{M_1^8M_5^8}\tilde{\beta}^2e^{-2\phi}\, .
\end{align}
In the gauge choice $M_y=M_xM_1^4M_5^4$ used in the main text, it greatly simplifies to
\begin{align}
    \phi'' &= \alpha^2 e^{2\phi} M_1^8 M_5^8 - \tilde{\beta}^2 e^{-2\phi}\,.
\end{align}
The Einstein equations (without fixing a gauge) are
\begin{align*}
        &(\mathbf{xx}): & \frac{L_x''}{L_x}
        -\frac{L_x'}{L_x}\frac{L_y'}{L_y}
        +4\frac{L_x'}{L_x}\left(\frac{L_1'}{L_1}+\frac{L_5'}{L_5}\right) &= 
        2\frac{L_x'}{L_x}\phi'
        - \frac{1}{4}e^{2\phi}\frac{L_y^2}{L_x^2}
        \left(\alpha^2+\frac{\tilde{\beta}^2 }{L_1^8 L_5^8}+\frac{\gamma^2 }{L_1^{8}} + \frac{\tilde{\delta}^2}{L_5^8}\right)\, ,\\
        &(\mathbf{yy}): & \frac{L_x''}{L_x}
        +4\left(\frac{L_1''}{L_1}+\frac{L_5''}{L_5}\right)
        -\frac{L_y'}{L_y}\left(\frac{L_x'}{L_x} + 4\frac{L_1'}{L_1}+4\frac{L_5'}{L_5}\right)
        &= 
        2\phi''
        -2\frac{L_y'}{L_y}\phi'
        + \frac{1}{4}e^{2\phi}\frac{L_y^2}{L_x^2}
        \left(\alpha^2+\frac{\tilde{\beta}^2 }{L_1^8 L_5^8}+\frac{\gamma^2 }{L_1^{8}} + \frac{\tilde{\delta}^2}{L_5^8}\right)\, ,\\
        &(\mathbf{xy}): & \alpha\tilde{\beta} + \gamma\tilde{\delta} &= 0\, ,\\
        &(\mathbb{T}_1): & \frac{L_1''}{L_1}
        +\frac{L_1'}{L_1}\left(\frac{L_x'}{L_x}-\frac{L_y'}{L_y}+4\frac{L_5'}{L_5}+3\frac{L_1'}{L_1}\right) &= 
        2\frac{L_1'}{L_1}\phi'
        + \frac{1}{4}e^{2\phi}\frac{L_y^2}{L_x^2}
        \left(\alpha^2-\frac{\tilde{\beta}^2 }{L_1^8 L_5^8}-\frac{\gamma^2 }{L_1^{8}} + \frac{\tilde{\delta}^2}{L_5^8}\right)\, ,\\
        &(\mathbb{T}_5): & \frac{L_5''}{L_5}
        +\frac{L_5'}{L_5}\left(\frac{L_x'}{L_x}-\frac{L_y'}{L_y}+4\frac{L_1'}{L_1}+3\frac{L_5'}{L_5}\right) &= 
        2\frac{L_5'}{L_5}\phi'
        + \frac{1}{4}e^{2\phi}\frac{L_y^2}{L_x^2}
        \left(\alpha^2-\frac{\tilde{\beta}^2 }{L_1^8 L_5^8}+\frac{\gamma^2 }{L_1^{8}} -\frac{\tilde{\delta}^2}{L_5^8}\right)\, .
\end{align*}
Upon gauge fixing $M_y=M_xM_1^4M_5^4$ (equivalently, $L_y = L_x L_1^4 L_5^4 e^{-2\phi}$), the $(\mathbf{xx})$, $(\mathbb{T}_1)$ and $(\mathbb{T}_5)$ components simplify considerably:
\begin{align}
        \frac{\dd}{\dd y} \left(\frac{L_x'}{L_x}\right)
        &=
        - \frac{1}{4}e^{-2\phi}
        \left(\alpha^2L_1^8 L_5^8+\tilde{\beta}^2+\gamma^2 L_5^8 + \tilde{\delta}^2 L_1^8 \right)\, ,\\
        \frac{\dd}{\dd y} \left(\frac{L_1'}{L_1}\right) &= 
        \frac{1}{4}e^{-2\phi}
        \left(\alpha^2 L_1^8 L_5^8 - \tilde{\beta}^2 -\gamma^2L_5^{8} + \tilde{\delta}^2L_1^8\right)\, ,\\
        \frac{\dd}{\dd y} \left(\frac{L_5'}{L_5}\right) &= 
        \frac{1}{4}e^{-2\phi}
        \left(\alpha^2 L_1^8 L_5^8 - \tilde{\beta}^2 +\gamma^2L_5^{8} - \tilde{\delta}^2L_1^8\right)\, ,
\end{align}
while the $(\mathbf{yy})$ component does not:
\begin{align*}
        \frac{L_x''}{L_x}
        +4\left(\frac{L_1''}{L_1}+\frac{L_5''}{L_5}\right)
        -\left(\frac{L_x'}{L_x} + 4\frac{L_1'}{L_1}+4\frac{L_5'}{L_5}-2\phi'\right)^2
        &= 
        2\phi''
        + \frac{1}{4}e^{-2\phi}
        \left(\alpha^2L_1^8 L_5^8+\tilde{\beta}^2+\gamma^2L_5^{8} + \tilde{\delta}^2L_1^8\right)\, .
\end{align*}

\paragraph{Killing spinor equations}\ \\
The Euclidean SUSY variations we use are taken from \cite{Aguilar-Gutierrez:2022kvk}:
\begin{align}
    \delta_\epsilon \lambda &= \frac{1}{2}\left(\partial_\mu \phi-i e^\phi\right) \Gamma^\mu F_\mu \epsilon \, ,\\
    \delta_\epsilon \psi_\mu &= D_\mu \epsilon+\frac{\mathrm{i}}{8} \mathrm{e}^\phi\left(\Gamma^\nu F_\nu+\frac{i^z}{2 \cdot 5!} \Gamma^{\nu_1 \ldots \nu_5} F_{\nu_1 \ldots \nu_5}\right) \Gamma_\mu \epsilon \, ,
\end{align}
where $z=0,1$ correspond to the supposed ambiguity in the Euclidean gravitino variations described in \cite{Aguilar-Gutierrez:2022kvk}. We leave this parameter free in this appendix, and it is argued in the main text that the correct choice is $z=0$. Note that the SUSY spinor $\epsilon$ has positive chirality by convention\footnote{In \cite{Aguilar-Gutierrez:2022kvk}, this was interpreted as a free choice of projector, but by consistency with the Killing spinor equations, it should be fixed.}:
\begin{align}
    \Gamma_{\bar{1} \ldots \bar{8} \bar{x} \bar{y}} \epsilon = i\epsilon \label{eq:positive_chirality_IIB}
\end{align}
For our Ansatz (\eqref{eq:ansatz_D-1D7_metric}-\eqref{eq:ansatz_D-1D7_F1}), the dilatino Killing spinor equation becomes:
\begin{align}
& \phi' = \eta_d e^{\phi} \frac{L_y}{L_x L_1^{4} L_5^{4}} \left(\alpha L_1^{4} L_5^{4} -\eta_d \tilde{\beta} \right) \label{eq:phi_KS}\\
& (\Gamma_{\Bar{x}} + i \eta_d \Gamma_{\Bar{y}}) \epsilon = 0\, , \label{eq:epsilon_projector_d}
\end{align}
where $\eta_d=\pm 1$ is a sign ambiguity. It corresponds to the choice of projector for the SUSY parameter, as visible in \eqref{eq:epsilon_projector_d}.\\

Moving on to the gravitino variations, it is first useful to note that the only non-vanishing components of the spin connection are 
\begin{align}
    \omega_{\bar{i} \bar{i} \bar{y}} &= \frac{L_i'}{L_i L_y} \quad i=x,1,5 \ (\text{no sum over $\bar{i}$})
\end{align}
With this in hand, there are four independent components of the gravitino equations, which we write as
\begin{align}
    \frac{1}{2} \frac{L_x'}{L_y} \Gamma_{\bar{x} \bar{y}}\epsilon+\frac{i}{8} e^\phi L_x \Sigma \Gamma_{\bar{x}} \epsilon &= 0\\
    \partial_y \epsilon+\frac{i}{8} e^\phi L_y \Sigma \Gamma_{\bar{y}} \epsilon &= 0\\
    \frac{1}{2} \frac{L_1'}{L_y} \Gamma_{\bar{i} \bar{y}}\epsilon+\frac{i}{8} e^\phi L_1 \Sigma \Gamma_{\bar{i}} \epsilon &= 0\, , \quad i=1,2,3,4\, ,\\
    \frac{1}{2} \frac{L_5'}{L_y} \Gamma_{\bar{j} \bar{y}}\epsilon+\frac{i}{8} e^\phi L_5 \Sigma \Gamma_{\bar{j}} \epsilon &= 0\, , \quad j=5,6,7,8\, ,
\end{align}
where $\Sigma$ is just shorthand for
\begin{align}
    \Sigma &= \Gamma^\nu F_\nu+\frac{i^z}{2 \cdot 5!} \Gamma^{\nu_1 \ldots \nu_5} F_{\nu_1 \ldots \nu_5} \nonumber \\
    &= \frac{\alpha}{L_x}\Gamma_{\bar{x}} + \frac{i\beta}{L_y}\Gamma_{\bar{y}} + \frac{i^z}{2}\frac{1}{L_1^4}\left(
    \frac{\gamma}{L_x} \left(
    \Gamma_{\bar{1}\bar{2}\bar{3}\bar{4}\bar{x}}
    + i\Gamma_{\bar{5}\bar{6}\bar{7}\bar{8}\bar{y}} \right)
    + \frac{i\delta}{L_y} \left(\Gamma_{\bar{1}\bar{2}\bar{3}\bar{4}\bar{y}}
    - i \Gamma_{\bar{5}\bar{6}\bar{7}\bar{8}\bar{x}} \right)
    \right)\, .
\end{align}
Making use of the projector from the dilatino variation \eqref{eq:epsilon_projector_d} and the positive chirality of $\epsilon$ \eqref{eq:positive_chirality_IIB}, we obtain
\begin{align}
    \Sigma \Gamma_{\bar{x}} \epsilon &= -i\eta_d \Sigma \Gamma_{\bar{y}} \epsilon = 
    \left(\frac{\alpha}{L_x} + \eta_d\frac{\beta}{L_y}\right)\epsilon + \frac{i^z}{2}\frac{1}{L_1^4}\left(
    \frac{\gamma}{L_x}
    + \eta_d\frac{\delta}{L_y} 
    \right)\left(
    \Gamma_{\bar{1}\bar{2}\bar{3}\bar{4}}
    + \eta_d\Gamma_{\bar{5}\bar{6}\bar{7}\bar{8}} \right) \epsilon
\end{align}
This in turn allows us to write the $x$ and $y$ components of the gravitino variations as
\begin{align}
    \partial_y \epsilon &= -\frac{1}{2} \frac{L_x'}{L_x} \epsilon = \eta_d \frac{1}{8} e^\phi L_y \Sigma \Gamma_{\bar{x}} \epsilon
\end{align}
Using the above, we can solve for the functional dependence of the SUSY spinor:
\begin{align}
    \epsilon(y) = \frac{\epsilon_0}{\sqrt{L_x(y)}}
\end{align}
The remaining three equations simplify to
\begin{align}
    \frac{L_x'}{L_x} \epsilon + \frac{1}{4} \eta_d  L_ye^\phi \left[\left(\frac{\alpha}{L_x} + \eta_d\frac{\beta}{L_y} \right) + \frac{i^z}{L_1^4}\left(
    \frac{\gamma}{L_x}
    + \eta_d\frac{\delta}{L_y} \right)\Gamma_{\bar{1}\bar{2}\bar{3}\bar{4}}\right] \epsilon&= 0\\
    \frac{L_1'}{L_1} \epsilon + \frac{1}{4} \eta_d L_y  e^\phi\left[-\left(\frac{\alpha}{L_x}-\eta_d \frac{\beta}{L_y}\right) + \frac{i^z}{L_1^4}\left(
    \frac{\gamma}{L_x} 
    - \eta_d\frac{\delta}{L_y} \right)\Gamma_{\bar{1}\bar{2}\bar{3}\bar{4}}\right]\epsilon &= 0\\
    \frac{L_5'}{L_5} \epsilon + \frac{1}{4} \eta_d L_y e^\phi \left[-\left(\frac{\alpha}{L_x}-\eta_d \frac{\beta}{L_y}\right) - \frac{i^z}{L_1^4}\left(
    \frac{\gamma}{L_x} 
    - \eta_d\frac{\delta}{L_y} \right)\Gamma_{\bar{1}\bar{2}\bar{3}\bar{4}}\right]\epsilon &= 0\, .
\end{align}
At this point, as in \cite{Aguilar-Gutierrez:2022kvk}, one has to impose a further projector equation on $\epsilon$, 
\begin{align}
    \Gamma_{\bar{1}\bar{2}\bar{3}\bar{4}}\epsilon = \eta_p \epsilon\, .
\end{align}
Again, $\eta_p=\pm 1$ is a sign ambiguity corresponding to the choice of projector. With this in hand, the remaining three Killing spinor equations are
\begin{align}
    \frac{L_x'}{L_x} + \frac{1}{4} \eta_d  \frac{L_y}{L_x L_1^{4} L_5^{4}} e^\phi \left[\left(\alpha L_1^{4} L_5^{4} + \eta_d \tilde{\beta} \right) + i^z\eta_p\left(
    \gamma L_5^{4}
    + \eta_d \tilde{\delta} L_1^{4} \right)\right] &= 0\\
    \frac{L_1'}{L_1} + \frac{1}{4} \eta_d  \frac{L_y}{L_x L_1^{4} L_5^{4}} e^\phi \left[-\left(\alpha L_1^{4} L_5^{4} - \eta_d \tilde{\beta} \right) + i^z\eta_p\left(
    \gamma L_5^{4}
    - \eta_d \tilde{\delta} L_1^{4} \right)\right] &= 0\\
    \frac{L_5'}{L_5} + \frac{1}{4} \eta_d  \frac{L_y}{L_x L_1^{4} L_5^{4}} e^\phi \left[-\left(\alpha L_1^{4} L_5^{4} - \eta_d \tilde{\beta} \right) - i^z\eta_p\left(
    \gamma L_5^{4}
    - \eta_d \tilde{\delta} L_1^{4} \right)\right] &= 0\, .
\end{align}

{\small
\bibliography{refs}}
\bibliographystyle{utphys}

\newpage
\end{document}